\documentclass[mathpazo]{cicp}
\usepackage{color}

\begin{document}
\title{New Approach for Error Reduction in the Volume Penalization Method}

\author[Iwakami W et.~al.]{Wakana Iwakami\affil{1}\comma\affil{2}\comma\corrauth,
Yuzuru Yatagai\affil{3}, Nozomu Hatakeyama\affil{4}, Yuji Hattori\affil{5}}
\address{
\affilnum{1}\ Yukawa Institute for Theoretical Physics, Kyoto University,
Oiwake-cho, Kitashirakawa, Sakyo-ku, Kyoto 606-8502, Japan \\
\affilnum{2}\ Advanced Research Institute for Science \& Engineering, Waseda University,
3-4-1 Okubo, Shinjuku, Tokyo 169-8555, Japan \\
\affilnum{3}\ Department of Applied Information Sciences, Graduate School of Information Sciences,
Tohoku University, 6-3-09 Aoba, Aramaki-aza, Aoba-ku, Sendai, Miyagi 980-8579, Japan\\
\affilnum{4}\ NICHe, Tohoku University, 6-6-10 Aoba, Aramaki-aza, Aoba-ku, Sendai, Miyagi 980-8579, Japan\\
\affilnum{5}\ Institute of Fluid Science, Tohoku University, 2-1-1 Katahira, Aoba-ku, Sendai, Miyagi, 980-8577, Japan\\}
\email{{\tt wakana@heap.phys.waseda.ac.jp} (W.~Iwakami)}


\begin{abstract}
A new approach for reducing error of the volume penalization method is proposed.
The mask function is modified by shifting the interface between solid and fluid by $\sqrt{\nu\eta}$ toward the fluid region, where $\nu$ and $\eta$ are the viscosity and the permeability, respectively.
The shift length $\sqrt{\nu\eta}$ is derived from the analytical solution of the one-dimensional diffusion equation with a penalization term.
The effect of the error reduction is verified numerically for the one-dimensional diffusion equation, Burgers' equation, and the two-dimensional Navier-Stokes equations.
The results show that 
the numerical error
is reduced except in the vicinity of the interface showing overall second-order accuracy, 
while it converges to a non-zero constant value  as the number of grid points increases for the original mask function.
However, the new approach is effective
when the grid resolution is sufficiently high so that the boundary layer, whose width is proportional to $\sqrt{\nu\eta}$, is resolved. 
Hence, the approach should be used when an appropriate combination of $\nu$ and $\eta$ is chosen with a given numerical grid.
\end{abstract}


\maketitle

\section{Introduction}
\label{sec1}

Flows around solid bodies have been investigated in a wide variety of fields in science and engineering.
Computational fluid dynamics has advantages in both visualizing flow fields and providing detailed data over experiments.
The flows around solid objects are often calculated using either a body-fitted grid system to impose boundary conditions or a set of appropriate orthogonal functions which satisfy the boundary conditions to expand the flow variables.
However, if there exist complex-shaped solid bodies or bodies which move or deform in the flow, it is not easy to generate a body-fitted grid system or to find a set of orthogonal functions; efficient computation is not possible at low cost by these methods.
The volume penalization (VP) method is one of the alternative methods to simulate flows in these complicated situations.

The VP method is one of the immersed boundary methods
which are classified into two types: {\it the continuous forcing approach} in which an external force term is added to a continuous equation of motion and {\it the discrete forcing approach} in which the force term is added to a discretized one\cite{mittal05}.
The VP method is the former type.
One 
can use it with the Fourier pseudo-spectral method;
many flows in which multiple solid bodies exist \cite{kevlahan01, kevlahan05, schneider05v1, schneider05v2}, the flows inside rigid boundaries \cite{schneider05v3, schneider08}, and the flows around moving bodies \cite{kolomenskiy09} have been simulated by the VP method.
Moreover, the VP method can be used with Chebyshev pseudo-spectral method, wavelet solvers, and other high-precision methods\cite{keetels07}.

In the VP method, a solid body is regarded as porous medium of low permeability.
There are two types of penalization modeling.
One is the $L^2$ penalization: the Navier-Stokes (N-S) equation is converted to the Darcy equation in the solid body;
and the other is the $H^1$ penalization: the N-S equation is transformed to the Brinkman equation in the solid body \cite{angot99, brinkman49}.
In the $L^2$ penalization, a damping force term which is called a penalization term and has a mask function $\chi$ and the permeability $\eta$ is added to the equation of motion.
Usually the step function, which is $0$ in the fluid region and $1$ in the solid region, is chosen as $\chi$.
The mask function activates the penalization term in the solid region so that the penalized N-S equation turns into the Darcy equation.

One of the advantages of the VP method is that there are rigorous results about convergence.
As permeability tends to zero, the penalized solution converges to the solution of the original (non penalized) problem with Dirichlet-type boundary conditions, e.g. no-slip boundary conditions.
Angot {\it et al.} proved mathematically that the upper bound for
the difference between the solutions of the original and penalized N-S equations, is $\mathrm{O}(\eta^{1/4})$ in the fluid region\cite{angot99}.
This upper bound is improved to $\mathrm{O}(\eta^{1/2})$ by Carbou and Fabrie\cite{carbou03}.
Kevlahan and Ghidaglia\cite{kevlahan01} considered a stokes flow over a flat plate whose dynamics is reduced to the one-dimensional diffusion equation and showed analytically that the error between the original and penalized solutions is $\mathrm{O}(\eta^{1/2})$ in the fluid region.
Recently, Kadoch {\it et al.} applied the VP method to problems with Neumann-type boundary conditions, e.g. no-flux conditions\cite{kadoch12}.
They draw the same conclusion as Carbou and Fabrie\cite{carbou03} in the convergence property.

These results suggest that 
in principle the error derived from the penalization term
can be smaller than the discretization error by choosing sufficiently small $\eta$ 
with a fine grid which can resolve the internal boundary layer.
However, we can not always choose a sufficiently small value for $\eta$ when an explicit method is used for time development.
The relation $\Delta t \le C \eta$ should be fulfilled in order to ensure numerical stability, where $\Delta t$ is the time step and $C$ is a constant which depends on the method of time integration.
In this paper a new approach for reducing 
the numerical error in the penalization method
is proposed, which is effective in the range of relatively large $\eta$ 
for a moderately small value of $\nu$
so that the VP method can be used with high accuracy even if the explicit method is used for time integration.
The results are verified for the problems of the one-dimensional (1D) diffusion equation, Burgers' equation, and the two-dimensional (2D) N-S equations.

This paper is organized as follows.
First, in Section 2 we mention the modification of the mask function on the basis of the analytical solution of 1D diffusion equation.
Next, we apply it to a 1D problem of Burgers' equation in Section 3, and to a 2D problem of N-S equation in Section 4.
Finally, we conclude in Section 5.

\section{Modification of Mask Function}
\label{sec2}

First, we derive {\it shift length} which is a parameter of changing the mask function to reduce 
the numerical error in the penalization method,
based on the analytical solution of the 1D diffusion equation with a penalization term.
Next, we confirm the effect of error reduction numerically by solving the 1D diffusion equation. 

\subsection{Analytical solution for 1D diffusion equation}
\label{sec2_1}

First we consider the initial value problem of the 1D diffusion equation
\begin{equation}
\frac{\partial \theta}{\partial t}=\nu\frac{\partial^2 \theta}{\partial x^2},
\label{eq_diff1dorg}
\end{equation}
where $\nu$ is the diffusion coefficient.
Given the initial value
\begin{equation}
\theta(0,x)=
         C_0 \ \sin{(k_n x)} \ \ \ \mathrm{in} \ \ |x| \le L,
\label{eq_diff1d_init}
\end{equation}
the exact solution can be written as
\begin{equation}
\theta_\mathrm{exact}(t,x)=\left\{\begin{array}{lll}
         0 & \mathrm{in} & L < x \le \infty \\
         C_0 \ e^{-\nu k_{n}^{2}t} \sin{(k_n x)} & \mathrm{in}  & |x| \le L \\
         0 & \mathrm{in} & -\infty < x \le -L \\
\end{array}
          \right.,
\label{eq_sol_diff1d3}
\end{equation}
where $C_0$
is the initial amplitude,
$k_{n} = \frac{n\pi}{L}$ ($n=1, 2, \cdot\cdot\cdot$) is the wavenumber of the 1D diffusion equation,
and $L$ is
the location of the interface between fluid and solid regions.
Here we assume $\theta=0$ in the solid region.

Next we consider the 1D diffusion equation with a penalization term 
\begin{equation}
\frac{\partial \theta_\eta}{\partial t}=\nu\frac{\partial^2 \theta_\eta}{\partial x^2}-\frac{\chi}{\eta}\theta_\eta,
\label{eq_diff1d}
\end{equation}
where $\eta$ is the permeability,
and the mask function $\chi$ is
\begin{equation}
\chi(x)=\left\{\begin{array}{lll}
        0 & \mathrm{in} & \Omega_f \\
        1 & \mathrm{in} & \Omega_s
        \end{array}
        \right.,
\label{eq_mask}
\end{equation}
where $\Omega_f$ and $\Omega_s$ correspond to the fluid region $|x|\le L$ and the solid region $L < |x|\le \infty$, respectively.
Given the initial value of Eq.~(\ref{eq_diff1d_init}), the exact solution of Eq.~(\ref{eq_diff1d}) is written in
\begin{equation}
\theta_{\eta \ \mathrm{exact}}(t,x)= \left\{\begin{array}{lll}
        C_+e^{-\nu k_{n}^{'2}t} e^{-\alpha x} & \mathrm{in}& L < x \le \infty\\
        C_0\ e^{-\nu k_{n}^{'2}t} \sin{(k_{n}^{'}x)} & \mathrm{in}&|x|\le L \\
        C_-e^{-\nu k_{n}^{'2}t} e^{\alpha x} &\mathrm{in}& -\infty \le x < -L \\
        \end{array}
        \right.,
\label{eq_sol_diff1d}
\end{equation}
where
$k'_{n}$ is the wavenumber of a solution for the 1D penalized diffusion equation,
$C_\pm$ and $\alpha$ are determined by the condition of $C^1$ continuity at $x=\pm L$,
\begin{equation}
\theta_{\eta \ \mathrm{exact}}(t,\pm L-0)=\theta_{\eta \ \mathrm{exact}}(t,\pm L+0),
\label{eq_c0}
\end{equation}
\begin{equation}
 \frac{\partial \theta_{\eta \ \mathrm{exact}}}{\partial x}(t,\pm L-0)
=\frac{\partial \theta_{\eta \ \mathrm{exact}}}{\partial x}(t,\pm L+0), 
\label{eq_c1}
\end{equation}
which give 
\begin{equation}
C_\pm = \pm C_0\ e^{\alpha L}\sin{(k'_n L)},
\label{eq_cpm}
\end{equation}
\begin{equation}
\frac{\sin{(k'_n L)}}{\cos{(k'_n L)}}=-\frac{k'_n}{\alpha}.
\label{eq_ak1}
\end{equation}
In addition, on substituting Eq.~(\ref{eq_sol_diff1d}) into Eq.~(\ref{eq_diff1d}), we obtain
\begin{equation}
\alpha^2 = \frac{1}{\nu\eta} - k^{'2}_n.
\label{eq_ak2}
\end{equation}
The constants $\alpha$ and $k'_n$ can be numerically obtained from (\ref{eq_ak1}) and (\ref{eq_ak2}). 
Substituting $\alpha$ and $k'_n$ into Eq.~(\ref{eq_sol_diff1d}),
the solution $\theta_{\eta \ \mathrm{exact}}$ in Eq.~(\ref{eq_diff1d}) is determined.

By comparing Eqs.~(\ref{eq_sol_diff1d}) and (\ref{eq_sol_diff1d3}),
we replace $L$ in Eq.~(\ref{eq_sol_diff1d}) by
\begin{equation}
L_\mathrm{VP} = L + \epsilon,
\label{eq_Lvp}
\end{equation}
so that the wavenumber $k'_n$ in Eq.~(\ref{eq_sol_diff1d}) coincides with $k_n$ in Eq.~(\ref{eq_sol_diff1d3}).
Then the relation (\ref{eq_ak1}) turns out to be
\begin{equation}
\frac{\sin{k_n L_\mathrm{VP}}}{\cos{k_n L_\mathrm{VP}}}=-\frac{k_n}{\alpha}.
\label{eq_ak3}
\end{equation}
Substituting Eq.~(\ref{eq_Lvp}) into Eq.~(\ref{eq_ak3}), we obtain
\begin{equation}
\frac{\sin{(k_n L)} \cos{(k_n \epsilon)} +\cos{(k_n L)} \sin{(k_n \epsilon})}{\cos{(k_n L)}\cos{(k_n \epsilon)} - \sin{(k_n L)}\sin{(k_n \epsilon)}}=-\frac{k_n}{\alpha},
\label{eq_ak4}
\end{equation}
Assuming $|k_n\epsilon| \ll 1$ and $\sqrt{\nu\eta} \ll 1/k_n$,
the {\it shift length} $\epsilon$ is obtained by Eqs.~(\ref{eq_ak2}) and (\ref{eq_ak4}) as
\begin{equation}
\epsilon \approx -\sqrt{\nu\eta}.
\label{eq_nueta}
\end{equation}
If the interface between $\chi=0$ and $1$ is located at $x=\pm L_\mathrm{VP} = \pm(L+\epsilon)$,
the penalized numerical solutions
have the same wavenumber as $k_n$
in the range of $-L_\mathrm{VP}\le x \le L_\mathrm{VP}$.
Based on the above result, we modify the mask function as follows
\begin{equation}
\chi(x)=\left\{\begin{array}{lll}
        0 & \mathrm{in} & \Omega_{f}' \\
        1 & \mathrm{in} & \Omega_{s}'
        \end{array}
        \right.,
\label{eq_mask2}
\end{equation}
where $\Omega'_f$ is $|x|\le L_\mathrm{VP}$ and $\Omega'_s$ is $L_\mathrm{VP}\le|x|\le \infty$.
In this paper, we call the mask function (\ref{eq_mask2}) as the {\it shifted mask function}.

\subsection{Numerical setups}
\label{sec2_2}

In this section we verify the effect of the shifted mask function on error reduction by numerical simulations of the 1D diffusion equation with a penalization term.
The equation (\ref{eq_diff1d}) is discretized with the fourth-order Runge-Kutta method in time and the fourth-order Pad\'{e}-type compact finite difference scheme in space\cite{lele92}.
The viscosity is set to $\nu = 0.1$, and 
the permeability is $\eta=10^{-5}-10^{-2}$.
Both the original mask function (\ref{eq_mask}) and the shifted one (\ref{eq_mask2}) 
are used.

The whole computational domain $\Omega=\Omega_f + \Omega_s=\Omega'_f + \Omega'_s$ covers the range $|x|<L_b$, where $L_b \approx 2\pi$ whose precise value depends on the grid number $N$ for the shifted mask function.
The original interfaces between $\Omega_f$ and $\Omega_s$ are located at $x=\pm L= \pm \pi$,
and the shifted interfaces between $\Omega'_f$ and $\Omega'_s$ are located at $x=\pm L_\mathrm{VP}= \pm(\pi - \sqrt{\nu\eta})$.

The values at the boundaries $x=\pm L_b$ of the computational domain are extrapolated from the values inside the computational domain.
However, since the solid regions $L<|x|<L_b$, where $\theta$ almost vanishes because of the penalization term, are sufficiently large, the boundary conditions at $x=\pm L_b$ hardly affect the results. 

The grid number is 
$N=90-2000$,
and the time step is fixed to $\Delta t = 10^{-5}$.
These values are chosen to satisfy three stability conditions, 
$u_\mathrm{max}\Delta t/\Delta x < C_c$, 
$\nu \Delta t/(\Delta x)^2 < C_d$, 
and $\Delta t < C\eta$, 
which arise from the nonlinear term, the diffusion term, and the penalization term, 
respectively. 
Here $u_\mathrm{max}$ is the maximum velocity which is zero for the diffusion equation 
but non-zero for the other equations under consideration.
The constants are set to $C_c=2.85/\sqrt{3}$, $C_d=2.9/6.0$, and $C_\eta=1$, where 
$C_c$ and $C_d$ are obtained by Lele \cite{lele92}.

Since the 1D diffusion equation is linear, it is sufficient to investigate a single 
wave solution.  
Thus the initial condition is set to Eq.~(\ref{eq_sol_diff1d3}) or Eq.~(\ref{eq_sol_diff1d}) with $t = 0$, $k'_n = k_n = 1$, and $C_0 = -1$. 

\subsection{Mask function on the discrete grid points}
\label{sec2_3}

\begin{figure}
\begin{center}
\includegraphics[width=7cm]{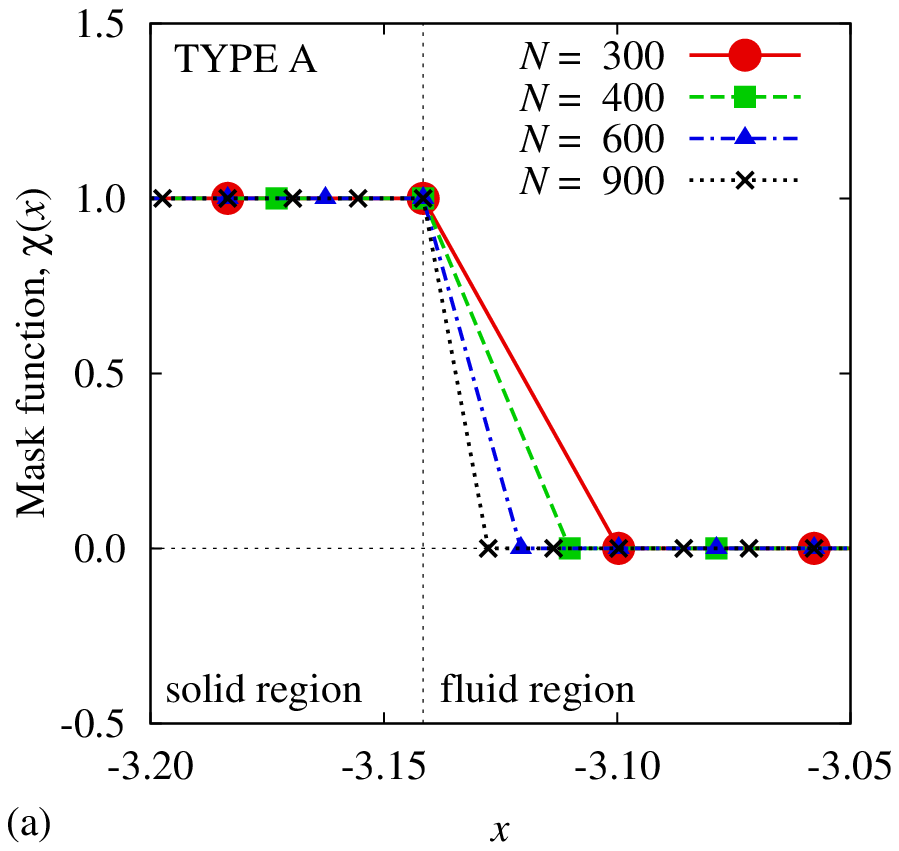}
\includegraphics[width=7cm]{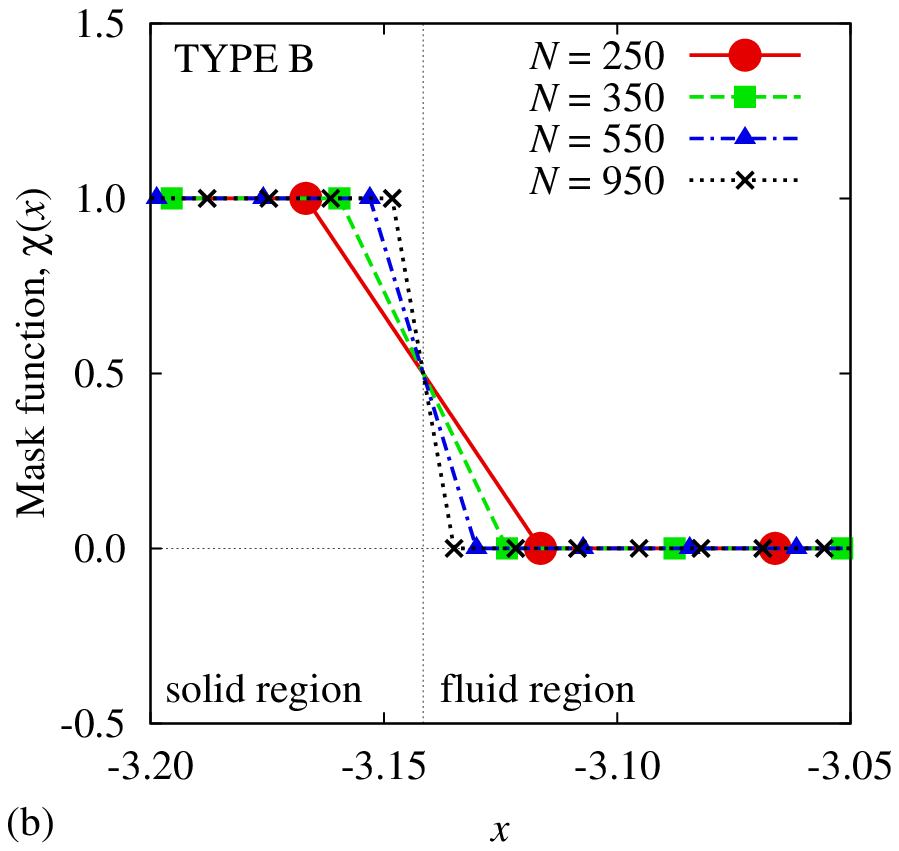}
\includegraphics[width=7cm]{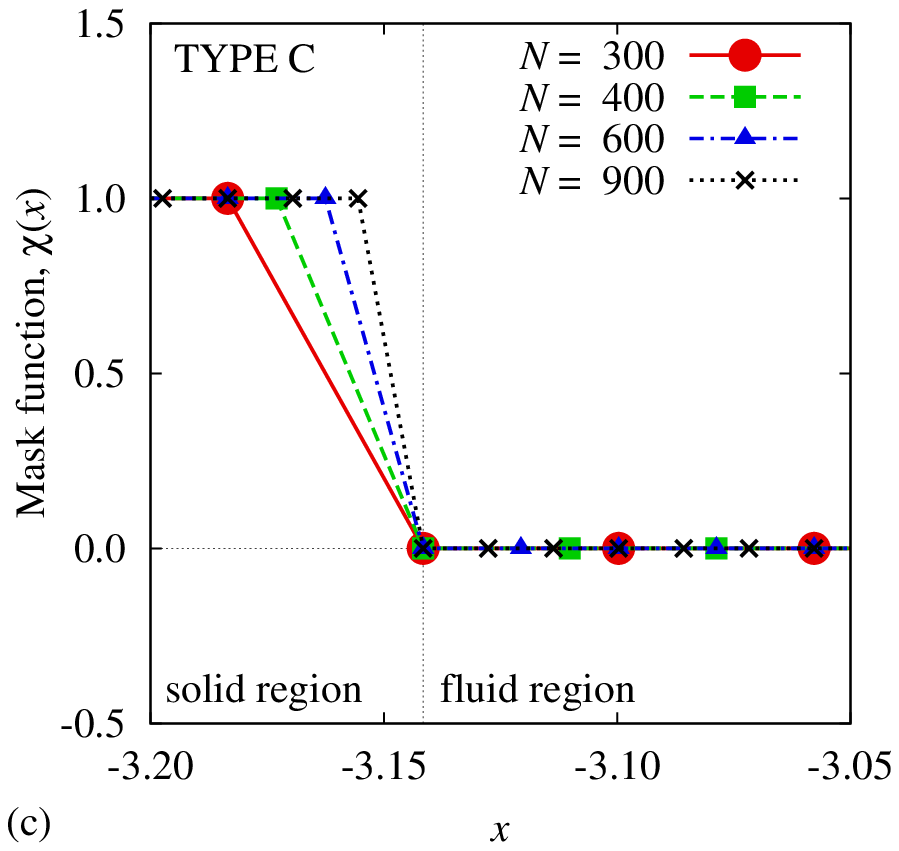}
\end{center}
\caption{
The original mask function
for (a) Type A, (b) Type B, and (c) Type C.
\label{fig_mask}}
\end{figure}

Here we consider a proper definition of the step function on discrete grid points.
We cannot express the step function rigorously on discrete grid points
since there is a non-zero gap between the grid points at which the value
jumps from $0$ to $1$.
Thus we should choose a mask function which gives the penalized numerical solution correctly converging to the penalized exact solution as $N$ tends to $\infty$.
Fig.~\ref{fig_mask} depicts three candidates for the original (non-shifted) mask functions near the boundary between $\Omega_s$ and $\Omega_f$. 
Focusing on the two points where $\chi$ jumps, 
the boundary coincides with the grid point where $\chi(\pm L) = 1$ for Type A (Fig.~\ref{fig_mask}a);
the boundary is located at the midpoint for Type B (Fig.~\ref{fig_mask} b);
the boundary coincides with the grid point where $\chi(\pm L) = 0$ for Type C (Fig.~\ref{fig_mask}c).

In order to find the most appropriate definition of the mask functions, we introduce an error defined as the root mean square of the difference between
the numerical solution and the exact penalized solution,
\begin{equation}
 \delta_{\eta \ \mathrm{exact}} \equiv \sqrt{\frac{\int_{\Omega_f} | \theta_\eta(t, x) - \theta_{\eta \ \mathrm{exact}} (t, x) |^{2}dx}{\int_{\Omega_f}\  dx}},
\label{eq_erretaexact}
\end{equation}
where $\theta_\eta$ is a numerical solution of Eq.~(\ref{eq_diff1d}), and $\theta_{\eta\ \mathrm{exact}}$ is given by Eq.~(\ref{eq_sol_diff1d}). 
In this case numerical calculation starts from $\theta_{\eta\ \mathrm{exact}}(0,x)$ of Eq.~(\ref{eq_sol_diff1d}).

\begin{figure}
\begin{center}
\includegraphics[width=7cm]{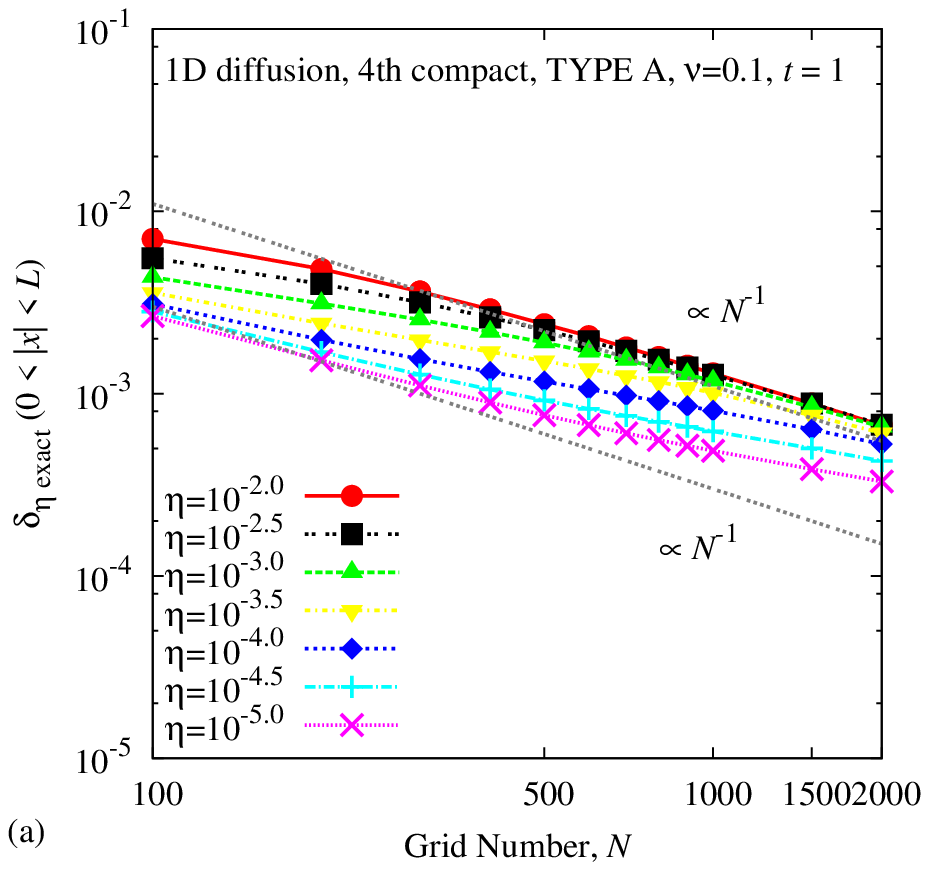}
\includegraphics[width=7cm]{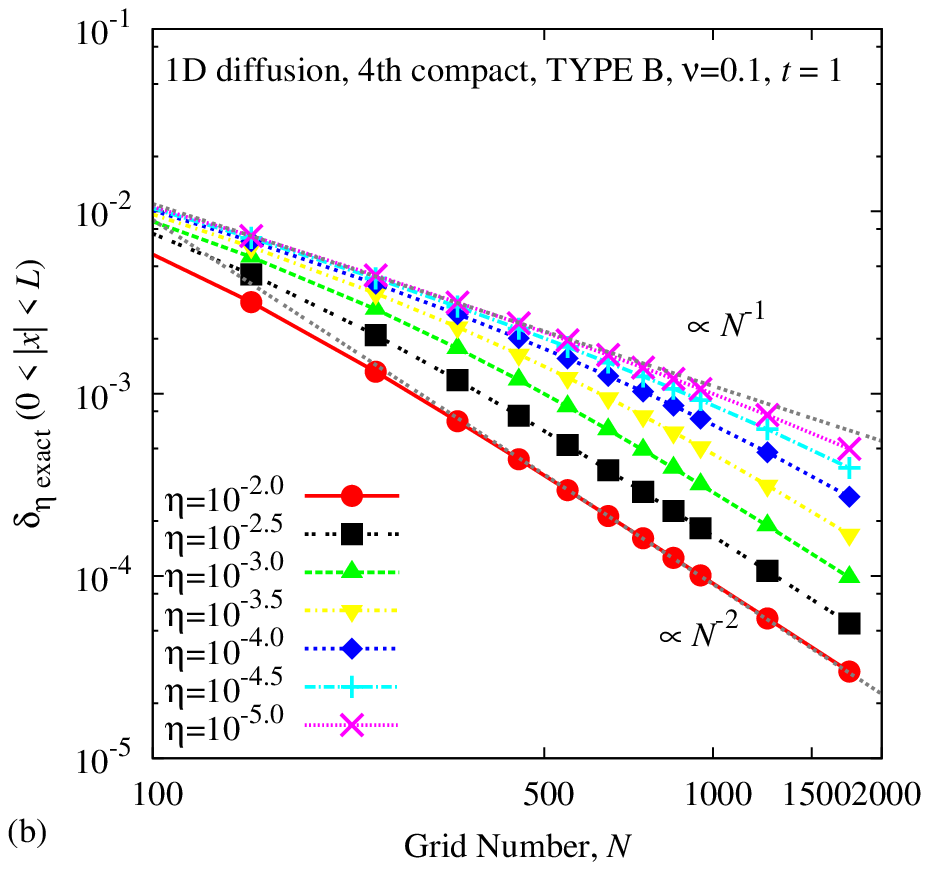}
\includegraphics[width=7cm]{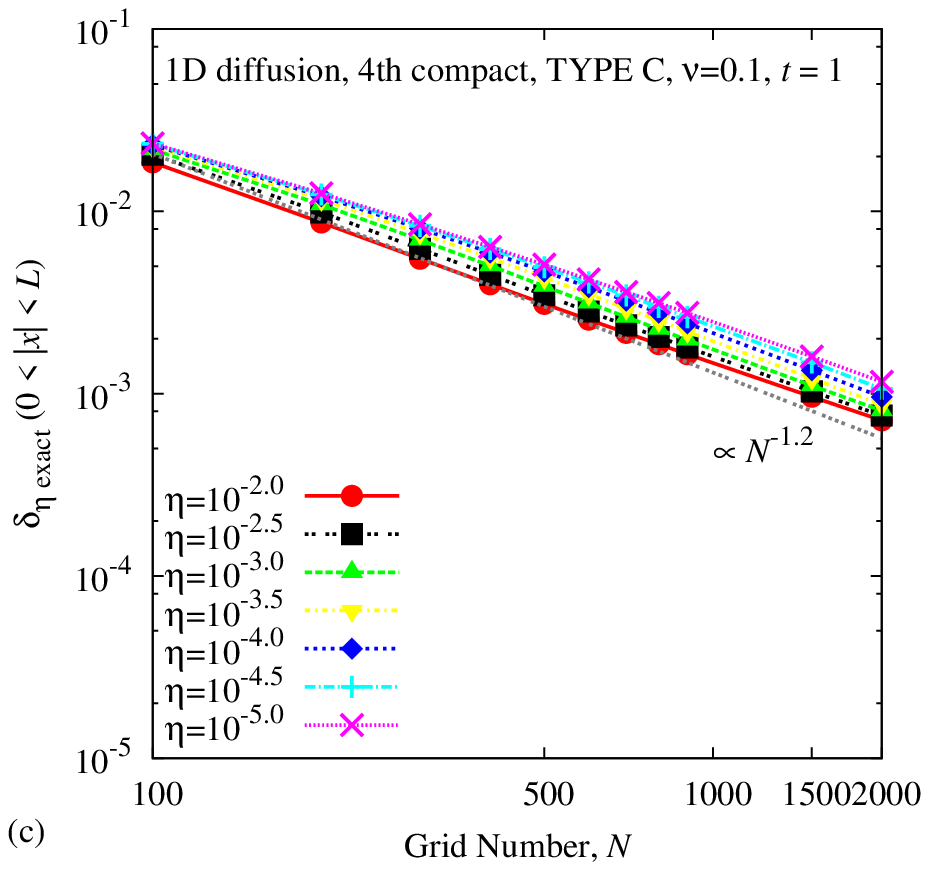}
\end{center}
\caption{
The error of the numerical solutions $\theta_\eta$ from the exact solutions $\theta_{\eta \ \mathrm{exact}}$ for the 1D penalized diffusion equation using the mask functions for (a) Type A, (b) Type B, and (c) Type C.
\label{fig_erretaexact}}
\end{figure}

The profiles of $\delta_{\eta \ \mathrm{exact}}$ as a function of $N$ are depicted in Fig.~\ref{fig_erretaexact}.
The error decreases with increasing $N$
for all three types of mask functions. 
Therefore,
the numerical solution with the original mask function monotonically converges to the exact solution of the penalized diffusion equation in Eq.~(\ref{eq_sol_diff1d}).
However, the convergence properties are different among the three types.
The error decreases in proportion to $N^{-1.0}$ for
Type~A, $N^{-2.0} \sim N^{-1.0}$ for Type~B, and $N^{-1.2}$ for Type~C.
Second-order accuracy achieved for Type~B and large $\eta$ is the highest accuracy.
Therefore, the best definition of the mask function is Type~B.

\begin{figure}[t]
\begin{center}
\includegraphics[width=7cm]{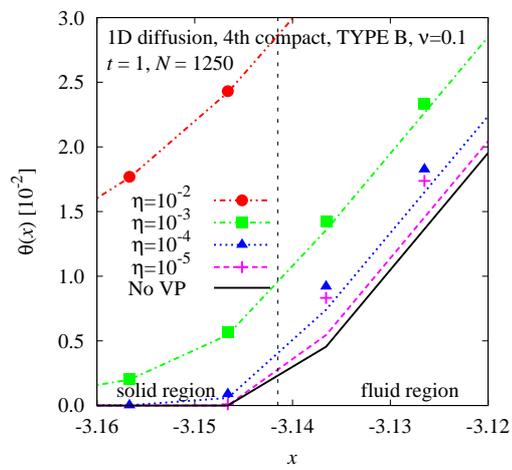}
\end{center}
\caption{
The numerical solutions $\theta_\eta$ for the 1D penalized diffusion equation using the original mask function for TYPE B
only near one interface between $\Omega_s$ and $\Omega_f$.
A vertical broken line indicates the interface.
The other broken lines denotes the exact solution $\theta_{\eta \ \mathrm{exact}}$ for the 1D penalized diffusion equation.
The exact solution $\theta_\mathrm{exact}$ for the 1D non-penalized diffusion equation is also shown with a solid line, termed as "No VP".
\label{fig_veldiff1d}}
\end{figure}

We also observe that $\delta_{\eta \ \mathrm{exact}}$ decreases with increasing $\eta$ for Type~B 
as resolution near the interfaces between $\Omega_s$ and $\Omega_f$ is insufficient for small $\eta$.
Fig.~\ref{fig_veldiff1d} shows the numerical and exact solutions of the penalized diffusion equation near one interface for Type B.
The solid line is the non-penalized exact solution $\theta_\mathrm{exact}$ (\ref{eq_sol_diff1d3}), 
the other lines are the penalized exact solution $\theta_{\eta\ \mathrm{exact}}$ (\ref{eq_sol_diff1d}), 
and
the symbols denote the penalized numerical solutions $\theta_\eta$.
The penalized solutions (broken lines) have leak from solid to fluid regions if we regard $\theta$ as flow velocity.
We call it a leaking area in a solid region.
The width of the leaking area is
\begin{equation}
1/\alpha \approx \sqrt{\nu\eta},
\label{eq_a}
\end{equation}
which is estimated by substituting Eq.~(\ref{eq_ak2}) into Eq.~(\ref{eq_sol_diff1d}) under the condition $\sqrt{\nu\eta} \ll 1$.
As the leaking area becomes smaller with decreasing $\eta$,
the penalized numerical solutions (symbols) deviate from the penalized exact solutions (broken lines),
while the penalized solutions approach the non-penalized one (solid line) with decreasing $\eta$.
As long as we use the same grid, the small leaking area cannot be resolved so that
$\delta_{\eta \ \mathrm{exact}}$ could be large for small $\eta$
(Fig.~\ref{fig_erretaexact}b).

We note that the error shows second-order accuracy although fourth-order schemes are used for the spatial discretization.
The reason is that the solution of the penalized diffusion equation in Eq.~(\ref{eq_diff1d}) has only $C^1$ continuity.
The second derivative of Eq.~(\ref{eq_sol_diff1d}) is discontinuous at the boundaries between $\Omega_f$ and $\Omega_s$.
Since the low-order accuracy is an intrinsic feature of the VP method,
we do not address further improvement of convergence property in this paper, though this problem is expected to be overcome in the future.

\subsection{Numerical results}
\label{sec2_4}

\begin{figure}[h]
\begin{center}
\includegraphics[width=10cm]{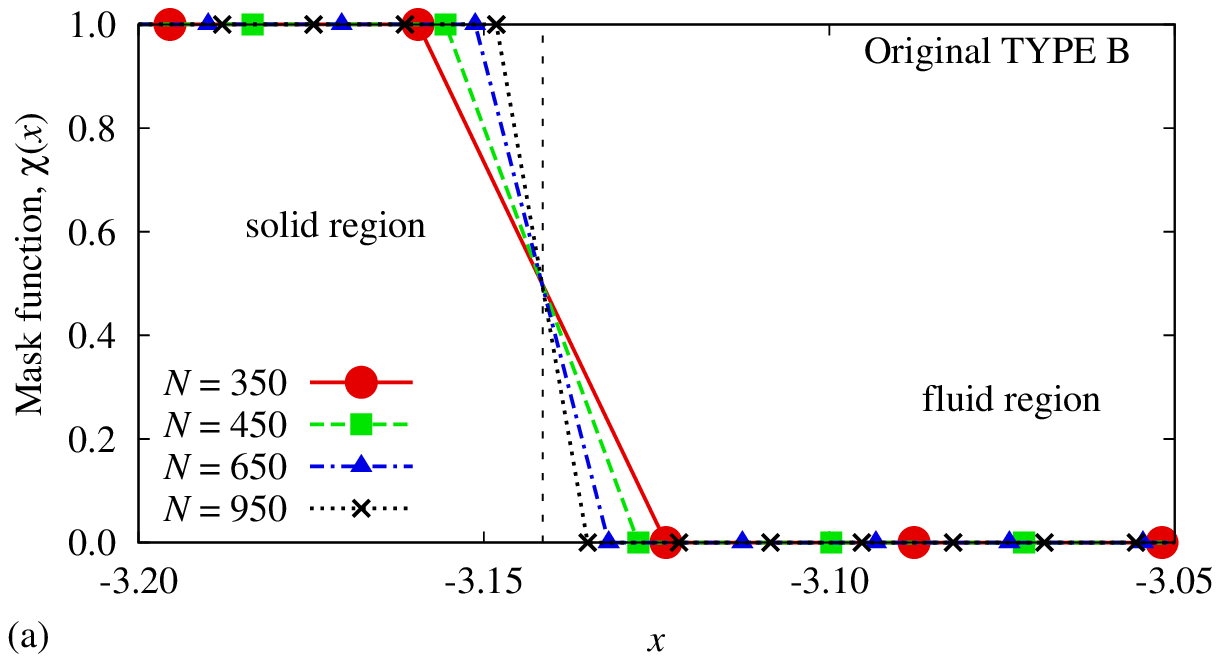}
\includegraphics[width=10cm]{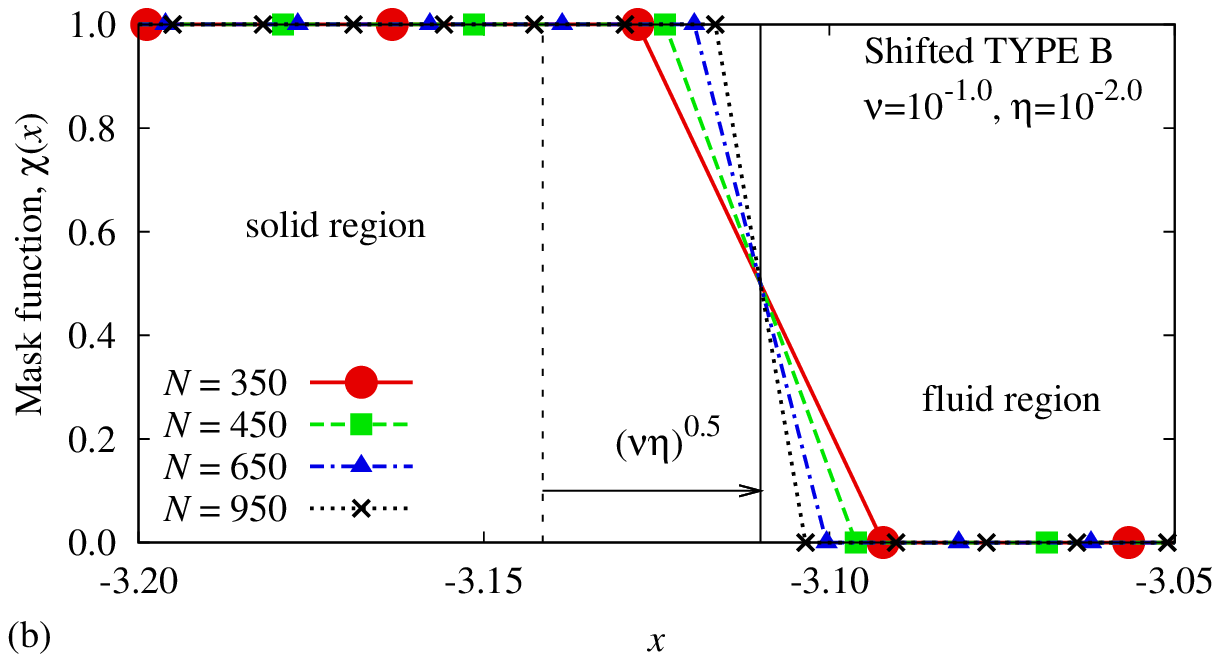}
\end{center}
\caption{
The original and shifted mask functions for Type B near the interface between $\Omega_s$ and $\Omega_f$.
The vertical thin broken line corresponds to the interface between $\Omega_s$ and $\Omega_f$,
and the vertical thin solid one indicate the shifted interface between $\Omega'_s$ and $\Omega'_f$.
\label{fig_mask2}}
\end{figure}

In this section we verify our method of error reduction.
The original and shifted mask functions for Type B near the solid boundary are shown in Fig.~\ref{fig_mask2}.
The original interfaces are located at $x=\pm L=\pm \pi$, shown by the thin broken lines.
In the shifted mask function the interface between fluid and solid is shifted toward the fluid region by $\sqrt{\nu\eta}$, which is obtained in Eq.~(\ref{eq_mask2}).
The shifted interface $x=-L_\mathrm{VP}=-(\pi - \sqrt{\nu\eta})$ is shown by the thin solid line in Fig.~\ref{fig_mask2} (b).
Since it is also located at the midpoint between two grid points,
the overall numerical domain $\Omega'$, which is $|x|\le L_b \approx 2\pi$, depends on $N$.

In order to verify the error reduction, we define a 
total error
as the root mean square of the difference between
the numerical solution and the exact solution,
\begin{equation}
 \delta_{\mathrm{tot}} \equiv \sqrt{\frac{\int_{\Omega_f} | \theta_\eta(t, x) - \theta_{\mathrm{exact}} (t, x) |^{2}dx}{\int_{\Omega_f}\  dx}},
\label{eq_errtot}
\end{equation}
where $\theta_\eta$ is the numerical solution of Eq.~(\ref{eq_diff1d}), and $\theta_\mathrm{exact}$ is the solution of the non-penalized 1D diffusion equation (\ref{eq_sol_diff1d3}). 
Simulation starts from $\theta_\mathrm{exact}(0,x)$ of Eq.~(\ref{eq_sol_diff1d3}).
The total error $\delta'_\mathrm{tot}$ for the shifted mask function
is defined by Eq.~(\ref{eq_errtot}), replacing $\Omega_f$ by $\Omega_f'$.
The total error may be expressed as
\begin{equation}
\delta_\mathrm{tot} = \delta_\eta + \delta_N + \delta_\mathrm{etc},
\label{eq_tot}
\end{equation}
where $\delta_\eta$, $\delta_N$, and $\delta_\mathrm{etc}$ are the 
error which is derived from the penalization term
and depends on $\eta$, the spatial discretization/truncation error which depends on $N$, and the sum of the other errors which is normally negligible, respectively.
The shifted mask function can decrease $\delta_\eta$ significantly, as discussed in the next paragraph.

\begin{figure}
\begin{center}
\includegraphics[width=7cm]{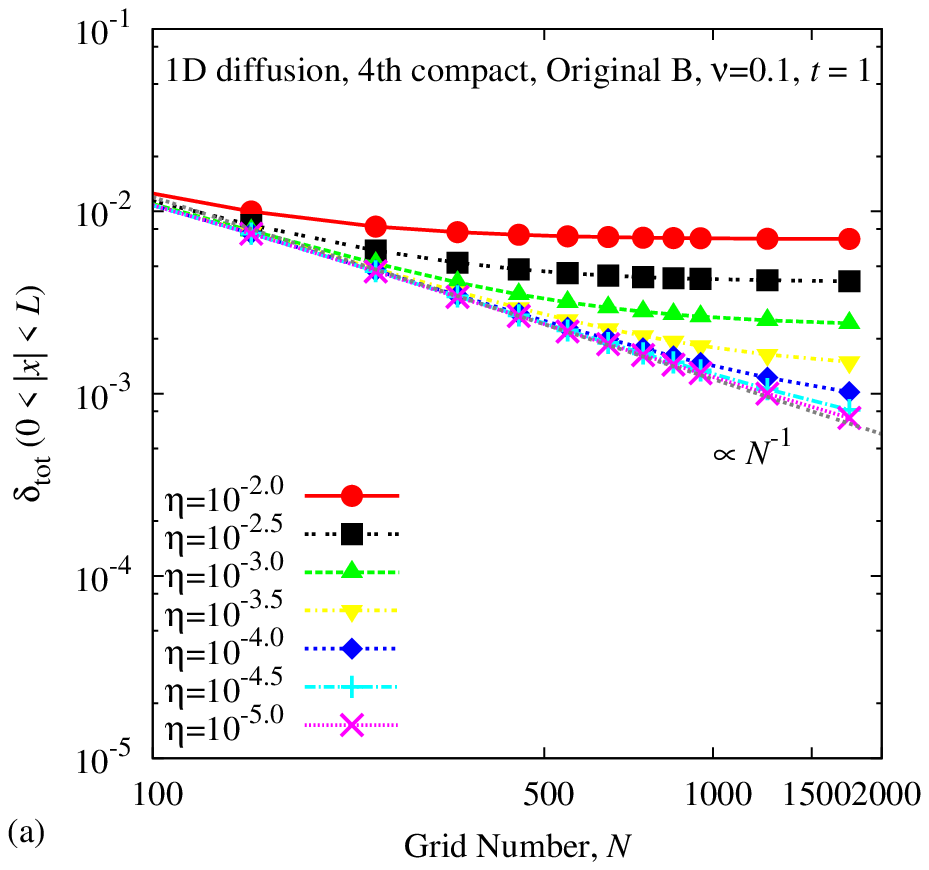}
\includegraphics[width=7cm]{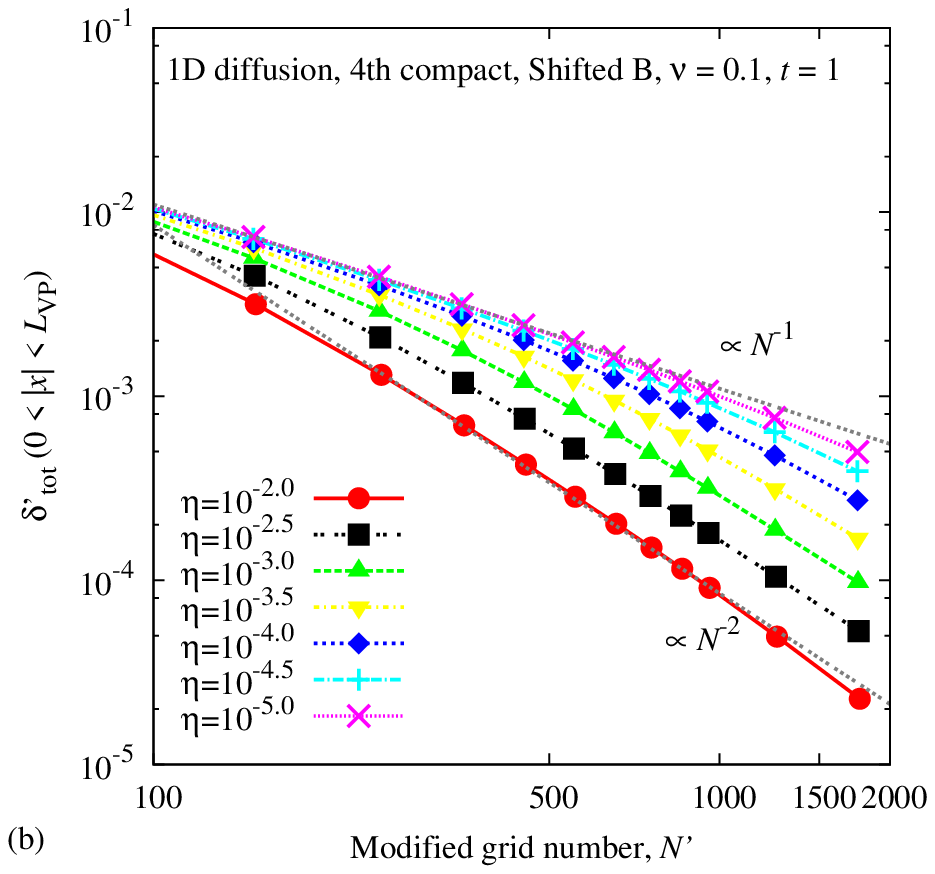}
\end{center}
\caption{
The total error $\delta_\mathrm{tot}$ and $\delta'_\mathrm{tot}$ as a function of $N$ and $N'$ for the 1D penalized diffusion equation with original and shifted mask functions for Type B, respectively.
\label{fig_errdiff1dB}}
\end{figure}

Fig.~\ref{fig_errdiff1dB} shows the grid number dependence of the total error for the numerical solutions of the penalized diffusion equation.
The results for the original and modified mask functions are shown in Fig.~\ref{fig_errdiff1dB} (a) and (b), respectively,
where we have modified the grid number as $N'=L/(L_b/N)$ taking account of the difference of $L_b$ among the simulations in Fig.~\ref{fig_errdiff1dB} (b).
When the resolution is low, the discretization error $\delta_{N}$ is dominant 
and the total error $\delta_\mathrm{tot}$ for the original mask function decreases as $N^{-1}$ 
(Fig.~\ref{fig_errdiff1dB}(a)).
If $\delta_\mathrm{tot}$ is on the line, the leaking area in the solid region cannot be resolved for a given $N$ as in the cases of $\eta=10^{-4}$ and $\eta=10^{-5}$ in Fig.~\ref{fig_veldiff1d}.
In the same region the total error $\delta'_\mathrm{tot}$ for the modified mask function hardly decreases, especially for smaller $\eta$
(Fig.~\ref{fig_errdiff1dB}(b)).
On the other hand, in the higher resolution, the total error $\delta_\mathrm{tot}$ for the original mask function approaches a constant with increasing $N$, while it decreases with $\eta$ as $O(\eta^{0.5})$.
The convergence property of $O(\eta^{0.5})$ agrees with the results by \cite{carbou03, kevlahan01}.
Hence the error derived from the penalization term $\delta_\eta$ would be dominant in the higher resolution (Fig.~\ref{fig_errdiff1dB}(a)).
In the same region the modified mask function decreases the total error $\delta'_\mathrm{tot}$ significantly for larger $\eta$ (Fig.~\ref{fig_errdiff1dB}(b)),
where the leaking area is accurately captured by the high-resolution grid.
The convergence of $\delta'_\mathrm{tot}$ is almost same as that of $\delta_{\eta \ \mathrm{exact}}$ in Fig.~\ref{fig_erretaexact} (b).
By modifying the mask function the total error decreases with increasing $N'$ 
instead of converging to a non-zero constant;
that is, the penalized numerical solutions with the shifted mask function converge to the non-penalized exact solutions.
Since the diffusion equation is linear
the Fourier modes of the solutions evolve independently.
Thus we can simply regard that each mode of wavenumber $k'_n$ is converted to the corresponding mode of wavenumber $k'_n$ by shifting the boundaries between fluid and solid regions by $\sqrt{\nu\eta}$.
Note that the shift length does not depend on the wavenumber. 

\begin{figure}
\begin{center}
\includegraphics[width=7cm]{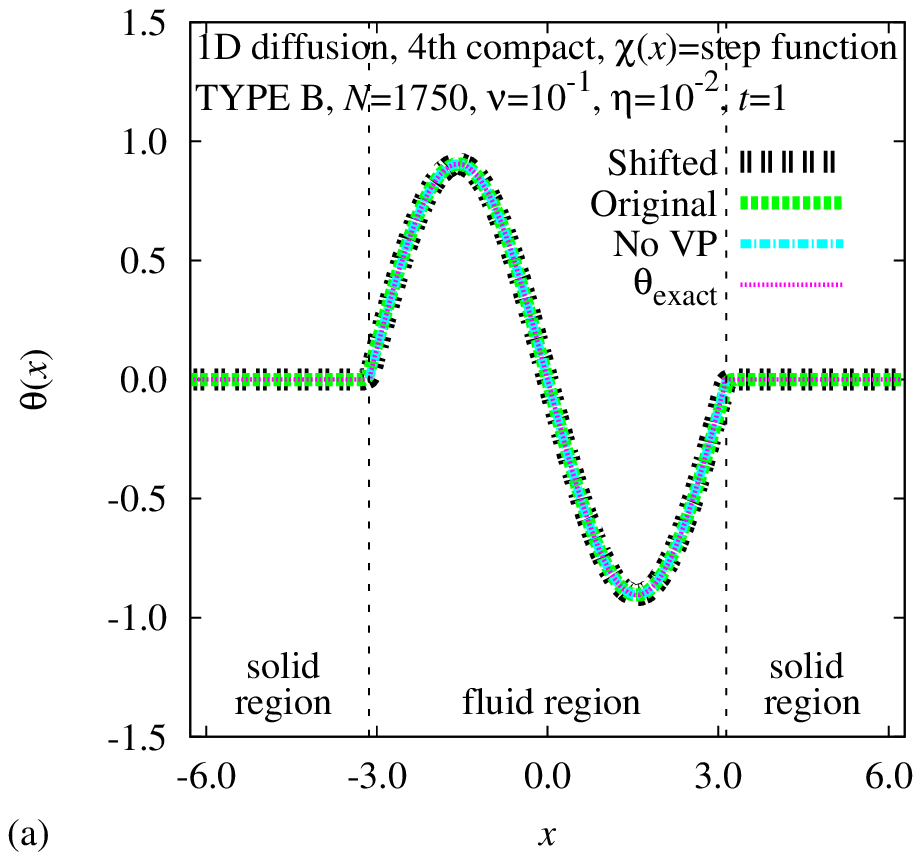}
\includegraphics[width=7cm]{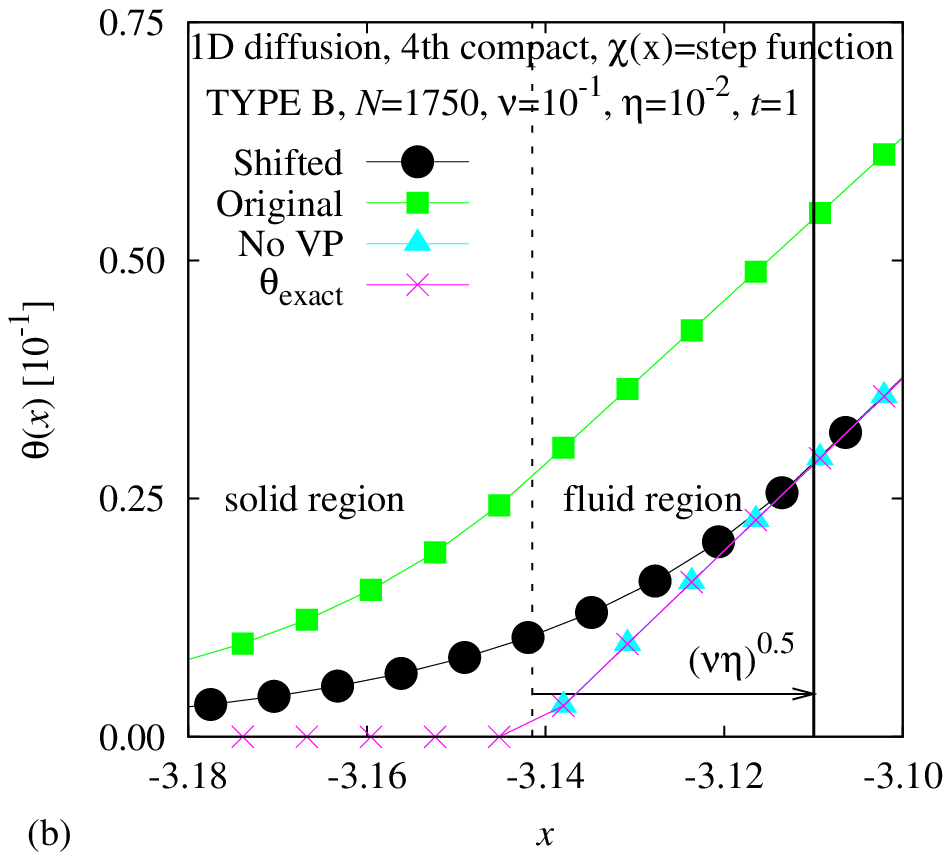}
\end{center}
\caption{
The comparison of the solutions for the 1D diffusion equation (a) in the whole region and (b) near the solid boundary.
\label{fig_veldiff1dB}}
\end{figure}

Finally we show why the shifted mask function works well. 
The solutions of the diffusion equation are shown in Fig.~\ref{fig_veldiff1dB}.
The vertical dashed lines are the original boundaries $\pm L$ between $\Omega_s$ and $\Omega_f$, and the vertical solid line is one of the shifted boundaries $\pm L_\mathrm{VP}$ between $\Omega'_s$ and $\Omega'_f$.
In Fig.~\ref{fig_veldiff1dB}~(a), 
the solutions are close to a sine wave in the fluid region and is approximately zero in the solid regions.
Both numerical solutions for original and shifted mask functions apparently agree with the exact solution of the non-penalized diffusion equation.
Close look at the boundary shown in Fig.~\ref{fig_veldiff1dB}~(b), however, reveals that 
the numerical solution for the original mask function deviates from the exact solution in the whole region.
On the other hand, the numerical solution for the shifted mask function agrees well with the exact solution for $-L_\mathrm{VP} < x < L_\mathrm{VP}$,
while there is a small difference between them in the vicinity of the interfaces $L_\mathrm{VP} < |x| < L$.
Although the small difference cannot be eliminated by our approach, the 
large deviation of the numerical solution from the exact solution
is confined to the area near the interfaces,
while the 
deviation prevails over
the entire domain for the original mask function.
If we use the original mask function, the difference of wave numbers between Eq.(\ref{eq_sol_diff1d3}) and Eq.(\ref{eq_sol_diff1d}) leads to large difference of the amplitudes between them as time proceeds owing to the factor of $e^{-\nu^2k_nt}$.
Thus the convergence of $\delta_\mathrm{tot}$ for $N$ at a point in $|x|<L$, except at $x=0$, has the same characteristics of approaching a constant with increasing $N$ as shown in Fig.~\ref{fig_errdiff1dB}~(a). 
On the other hand, the shifted mask function modifies the wavenumber of the numerical solution to coincide with the wavenumber of the exact solution.
Thus the shifted mask function can eliminate not only the difference of phases but also that of amplitudes
in $|x| \le L_\mathrm{VP}$.
Therefore, the confinement of the 
deviation from the exact solution
into the small region
near the interfaces 
has a striking effect on improving the accuracy of the VP method.

\section{Application to a Non-linear Equation}
\label{sec3}

\subsection{1D Burgers' equation \label{sec3_1}}

1D Burgers' equation,
\begin{equation}
\frac{\partial u}{\partial t}+u \frac{\partial u}{\partial x}=\nu\frac{\partial^2 u}{\partial x^2},
\label{eq_burg1d}
\end{equation}
can be regarded as the 1D compressible N-S equation without pressure\cite{burgers39}.
The 1D diffusion equation considered in the previous section 
is desirable for validating numerical methods
since both the original equation and the penalized equation have analytical solutions.
However, nonlinear effects cannot be addressed since it is linear.
Thus as the next problem we choose 1D Burgers' equation, which the simplest nonlinear equation related with the N-S equations and has analytical solutions.

Exact solutions of Eq.~(\ref{eq_burg1d}) are obtained using known formulae\cite{hopf50, cole51}
\begin{equation}
u_\mathrm{exact}(t,x) = 2\nu\frac{\theta_x}{\theta} = 2\nu\frac{\sum^{\infty}_{n=1}(n\pi/L)\exp[-\nu n^2 \pi^2 t/L^2]A_n \sin(n\pi x/L)}{A_0 + \sum^{\infty}_{n=1}\exp[-\nu n^2 \pi^2 t/L^2] A_n \cos(n\pi x/L)},
\label{eq_sol_burg1da}
\end{equation}
\begin{equation}
A_0 = \frac{1}{2L}\int^{L}_{-L}\theta_0(x)dx,
\label{eq_sol_burg1da_a0}
\end{equation}
\begin{equation}
A_n = \frac{1}{L}\int^{L}_{-L}\theta_0(x)\cos\frac{n\pi x}{L}dx,
\label{eq_sol_burg1da_an}
\end{equation}
\begin{equation}
\theta_0 (x) = C \exp\left(-\frac{1}{2\nu}\int^{x}_{0}u_0(\xi)d\xi\right),
\label{eq_sol_burg1da4}
\end{equation}
under the boundary condition
\begin{equation}
u(t,\pm L)=0,
\label{eq_burg1da_bc}
\end{equation}
where $\theta$ is the solution of the 1D diffusion equation (\ref{eq_diff1dorg}), $\theta_x=\frac{\partial \theta}{\partial x}$, and $C$ is an integral constant which is irrelevant to Eq.~(\ref{eq_sol_burg1da}).
In the following the initial condition is set as 
\begin{equation}
u(0,x)=u_0(x)= C_0 \sin (k_n x). 
\label{eq_burg1da_ic}
\end{equation}
In numerical calculation $u_\mathrm{exact}$ is obtained approximately.
The sum of Fourier modes are truncated at $n=100$, and the domain of integration $-L\le x \le L$ is divided into segments of width $\Delta x = 10^{-4}$. 
We find that the exact solution can be correctly obtained for $\nu = 0.1$.

In this section, we numerically solve 1D penalized Burgers' equation
\begin{equation}
\frac{\partial u_\eta}{\partial t}+u_\eta \frac{\partial u_\eta}{\partial x}=\nu\frac{\partial^2 u_\eta}{\partial x^2} - \frac{\chi}{\eta}u_\eta,
\label{eq_burg1da}
\end{equation}
where $\nu$ denotes the diffusion coefficient, and $\eta$ is the permeability.
Numerical setups are basically the same as those in the previous problem of the 1D diffusion equation (Section \ref{sec2_2}).
The parameters are set to $\nu = 0.1$, $\eta=10^{-5}-10^{-2}$, $C_0 = -1$, $k_n = 1$, $L = \pi$, $L_b \approx 2\pi$, $L_\mathrm{VP}=\pi-\sqrt{\nu\eta}$, $N=90-2000$, and $\Delta t = 10^{-5}$.
The original mask function (\ref{eq_mask})
and the shifted mask function (\ref{eq_mask2}) are also used.
The initial condition is given by Eq.~(\ref{eq_burg1da_ic}) in the fluid region, and $u(0, x)=0$ in the solid regions.

\subsection{Numerical results}

We verify whether the shifted mask function proposed in the Section \ref{sec2} is also effective for the nonlinear Burgers' equation.

\begin{figure}
\begin{center}
\includegraphics[width=7cm]{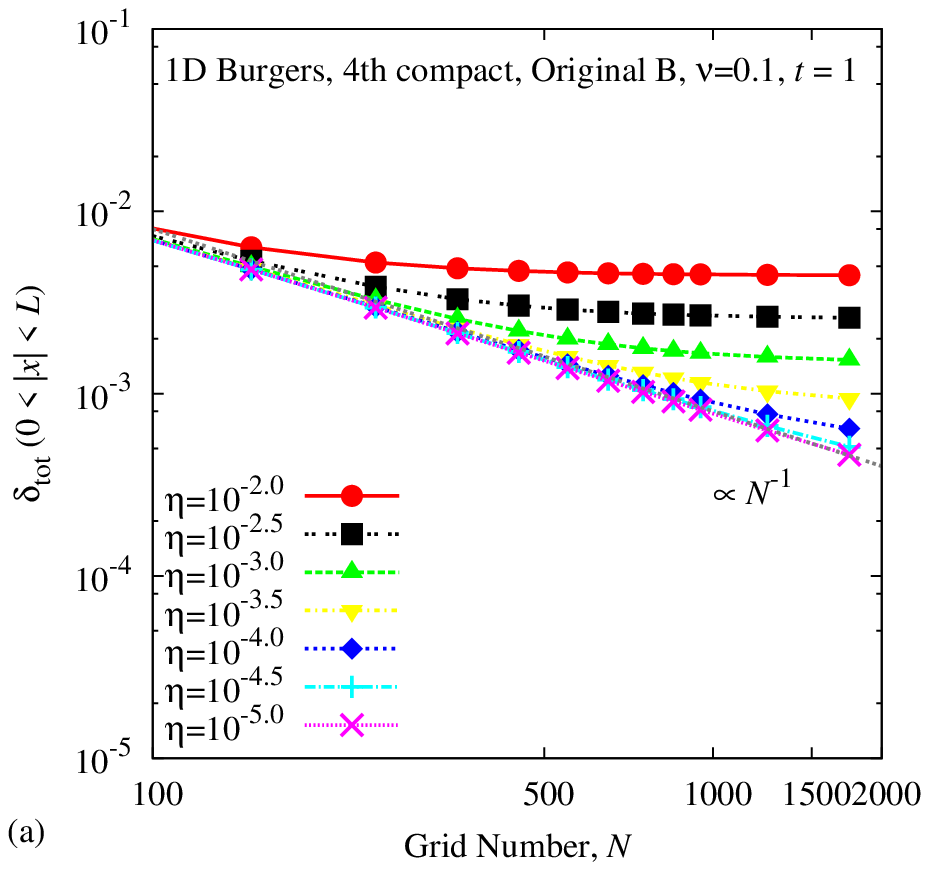}
\includegraphics[width=7cm]{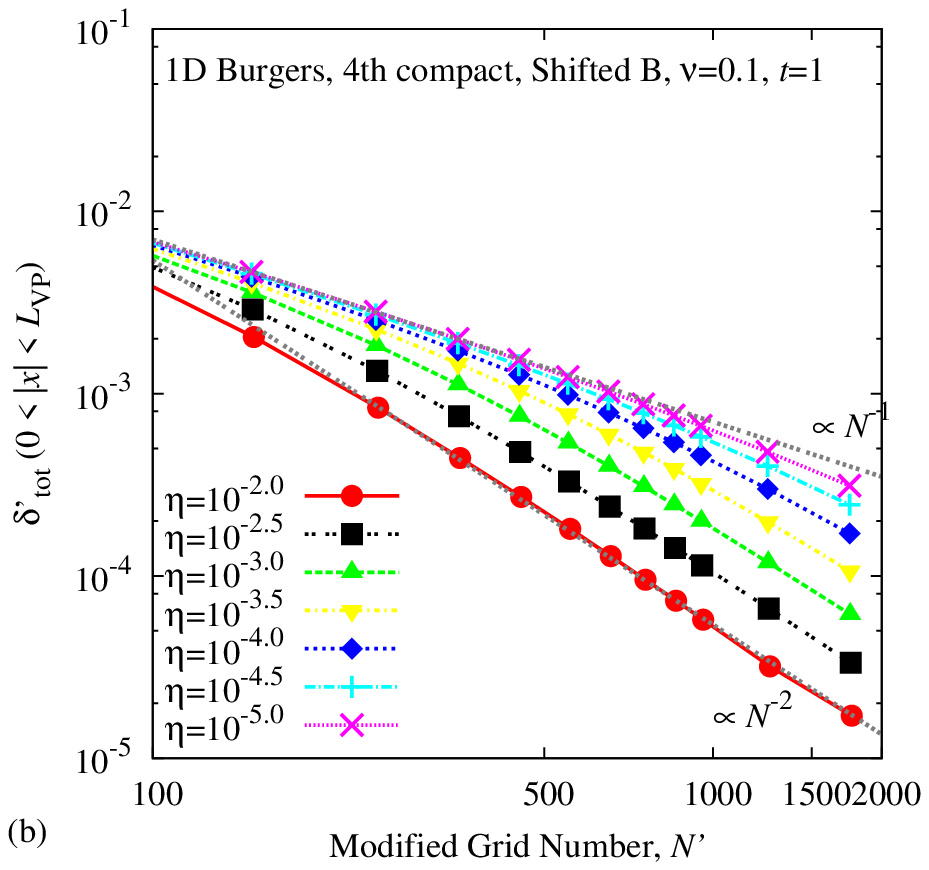}
\end{center}
\caption{
The total error $\delta_\mathrm{tot}$ and $\delta'_\mathrm{tot}$ as a function of $N$ and $N'$ for the 1D penalized Burgers' equation with original and shifted mask functions for Type B, respectively.
\label{fig_errburg1dB}}
\end{figure}

Fig.~\ref{fig_errburg1dB} shows the total error as a function of grid points.
In this section the total error $\delta_\mathrm{tot}$ 
is defined by
\begin{equation}
 \delta_{\mathrm{tot}} \equiv \sqrt{\frac{\int_{\Omega_f} | u_\eta(t, x) - u_{\mathrm{exact}} (t, x) |^{2}dx}{\int_{\Omega_f}\  dx}},
\label{eq_errtot2}
\end{equation}
where $u_\eta$ is the numerical solution (\ref{eq_sol_burg1da}), and $u_\mathrm{exact}$ is the solution of non-penalized Burgers' equation (\ref{eq_burg1d}). 
For the shifted mask function, $\delta'_\mathrm{tot}$ is defined by Eq.(\ref{eq_errtot2}) replacing $\Omega_f$ by $\Omega'_f$.
The convergence of $\delta_\mathrm{tot}$ and $\delta'_\mathrm{tot}$ for 1D Burgers' equation has the same characteristics as that for the diffusion equation shown in Fig.~\ref{fig_errdiff1dB}.
The penalized numerical solutions obtained by using the shifted mask functions converge to the exact solutions of original Burgers' equation for $-L_\mathrm{VP} <  x < L_\mathrm{VP}$.
Compared with the case of the original mask functions (Fig.~\ref{fig_errburg1dB}a), the total error for the shifted mask functions is reduced for large $N$, especially for large $\eta$ (Fig.~\ref{fig_errburg1dB}b).

\begin{figure}
\begin{center}
\includegraphics[width=7cm]{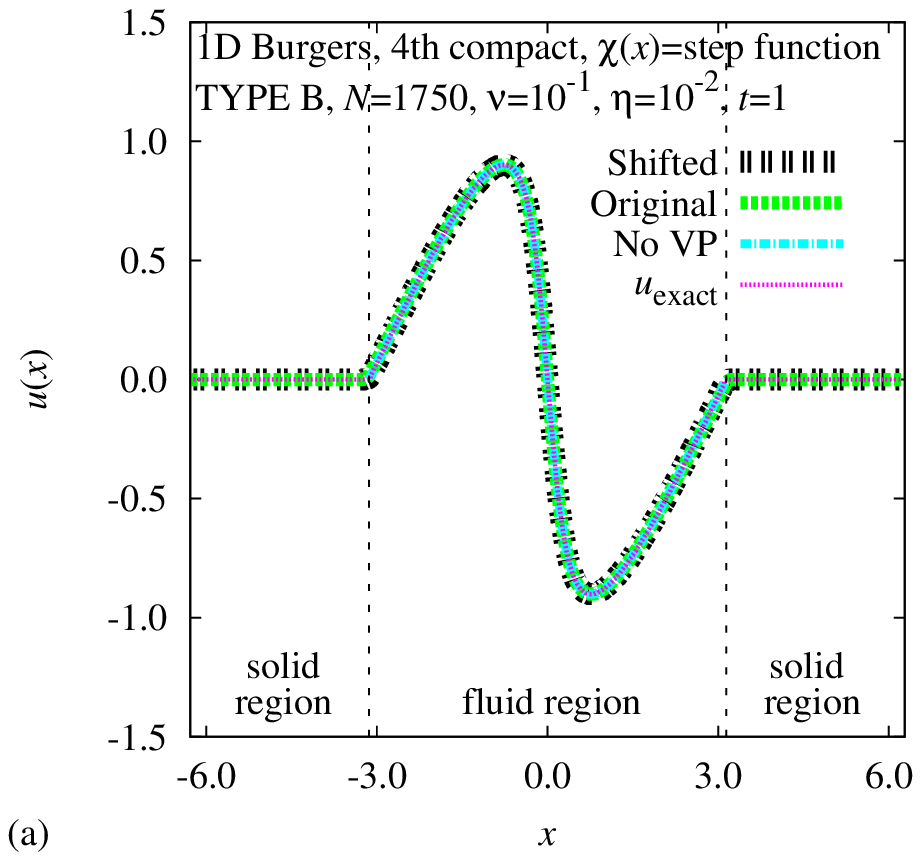}
\includegraphics[width=7cm]{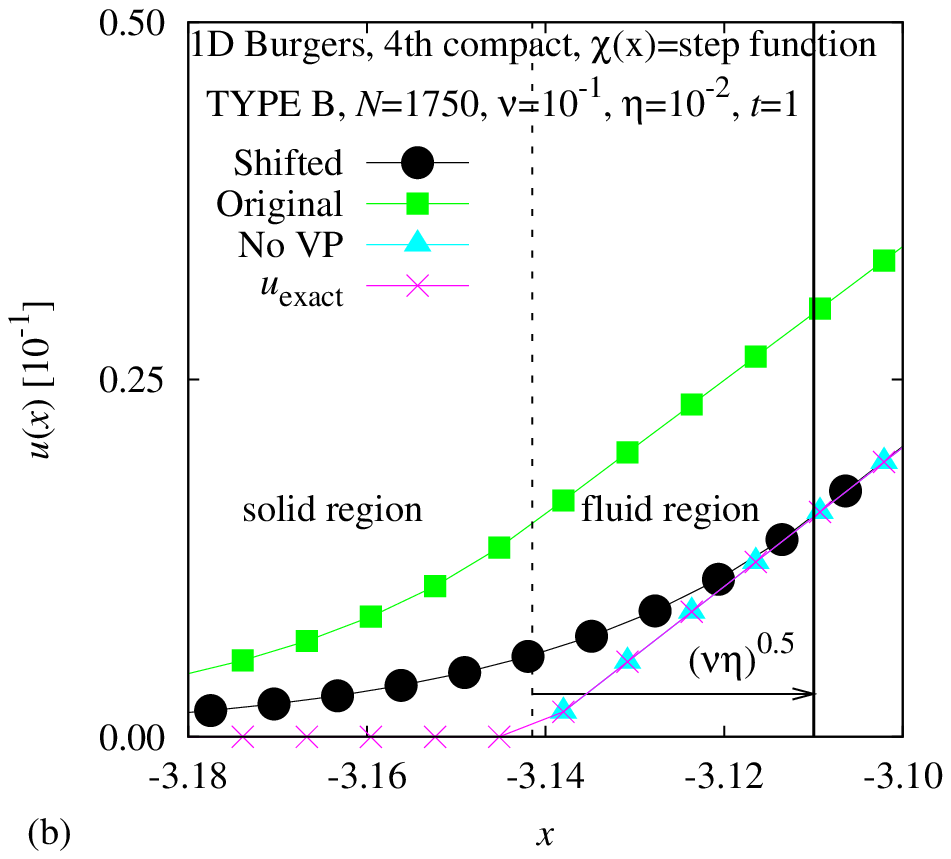}
\end{center}
\caption{
The comparison of the solutions for the 1D Burgers' equation (a) in the whole region and (b) near the solid boundary.
\label{fig_velburg1dB}}
\end{figure}

The solutions of 1D Burgers' equation in the whole region is shown in Fig.~\ref{fig_velburg1dB} (a).
The slope of $u$ at $x=0$ is steeper than that of $\theta$ in Fig.~\ref{fig_veldiff1dB} (a). 
The nonlinear term of 1D Burgers' equation makes the Fourier modes interact with each other exciting high wavenumber modes.
Nevertheless, the new technique of modifying the mask function gives the same effect on the numerical solutions of Burgers' equation.
Fig.~\ref{fig_velburg1dB} (b) confirms that the relation between the exact and numerical solutions for original and shifted mask functions near the interfaces between fluid and solid is also the same as that for the diffusion equation shown in Fig.~\ref{fig_veldiff1dB} (b).
The deviation of the numerical solution from the exact solution
is confined to be the vicinity of the interface.

\section{Application to 2D Problem}
\label{sec4}

Finally, we apply the method of reducing error to the 2D N-S equations and investigate whether it is effective.

\subsection{Taylor-Couette flow\label{sec4_1}}
The governing equations are the incompressible 2D N-S equations
\begin{equation}
\frac{\partial \mathbf{u}}{\partial t}+\mathbf{u} \cdot\nabla\mathbf{u}=-\frac{1}{\rho_0}\nabla p+\nu\nabla^2 \mathbf{u},
\label{eq_ns2d}
\end{equation}
\begin{equation}
\nabla\cdot\mathbf{u} = 0,
\end{equation}
where $\mathbf{u}$
is flow velocity,
$p$ is pressure, $\rho_0$ is density which is constant.
We consider the Taylor-Couette flow between two co-axial cylinders,
which is a steady solution of 2D N-S equations.
The inner and outer cylinders have the radii $R_1$ and $R_2$, and they rotate with the angular velocities $\omega_1$ and $\omega_2$, respectively.
The polar coordinate system $(r, \theta)$ is used in the following. 
The azimuthal component of velocity $u_{\theta \ \mathrm{exact}}$ is
\begin{equation}
u_{\theta \ \mathrm{exact}}(r)=\left\{\begin{array}{lll}
        r \omega_1 & \mathrm{in} & \Omega_{s1} \\
\displaystyle
        \frac{\omega_2 R_2^2 - \omega_1 R_1^2}{R_2^2 - R_1^2}r
       +\frac{(\omega_1-\omega_2) R^2_1 R^2_2}{R_2^2 - R_1^2}\frac{1}{r}
      & \mathrm{in} & \Omega_f \\
        r \omega_2 & \mathrm{in} & \Omega_{s2}
        \end{array}
        \right.,
\label{eq_sol_ns2d}
\end{equation}
while the radial component is $u_{r \ \mathrm{exact}}(r)=0$ in the whole domain\cite{tritton88}.
The overall computational domain is $\Omega =\{(x, y) | -\pi \le x, y \le \pi \}$ 
since the Fourier spectral method is used under doubly periodic boundary conditions.
The fluid region $\Omega_f$ is defined as $R_1 \le r \le R_2$.
The solid region $\Omega_s$ consists of the inner solid region $\Omega_{s1} = \{(x, y)\ |\ 0 \le r < R_1\}$ and the outer solid region $\Omega_{s2} = \{(x, y)\ |\ R_2 < r \  \mathrm{in} \  \Omega \}$.
Note that the fluid region is isolated as it is completely contained inside the periodic box 
and does not interact with the other fluid regions. 

\begin{figure}
\begin{center}
\includegraphics[width=10cm]{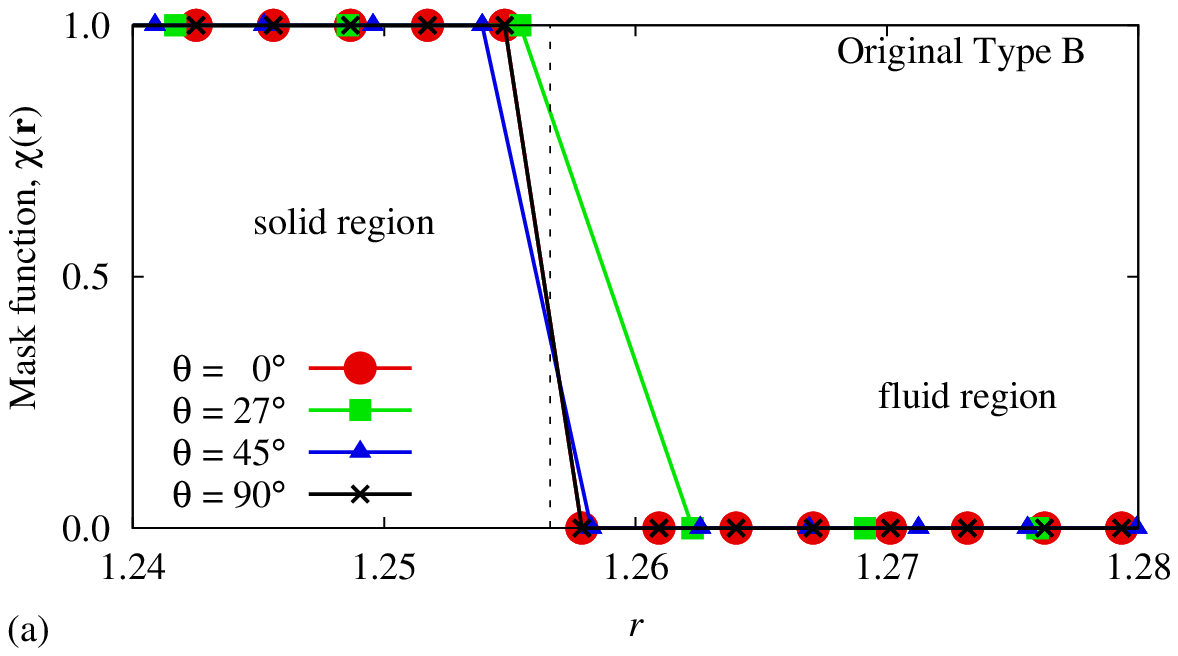}
\includegraphics[width=10cm]{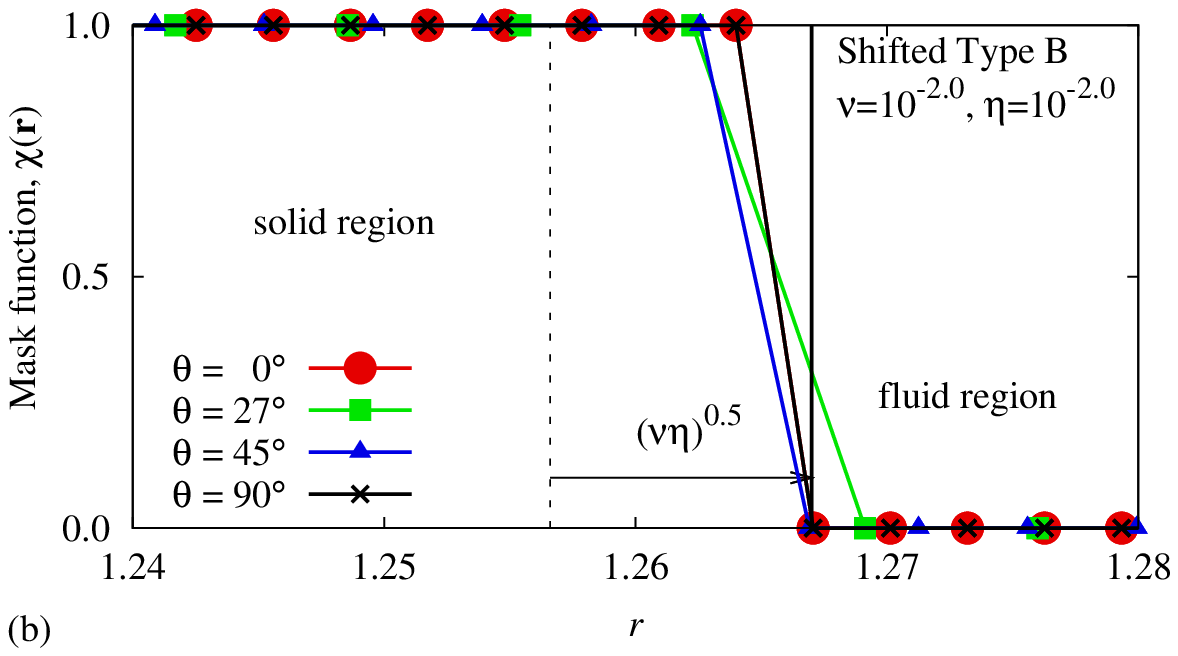}
\end{center}
\caption{
The distributions of $\chi(x)$
near the inner cylinder for the 2D problem:
(a) the original mask functions, (b) the shifted mask functions.
\label{fig_mask3}}
\end{figure}

We numerically solve the 2D N-S equations with a penalization term
\begin{equation}
\frac{\partial \mathbf{u}_\eta}{\partial t}+\mathbf{u}_\eta \cdot\nabla\mathbf{u}_\eta=-\frac{1}{\rho_0}\nabla p+\nu\nabla^2 \mathbf{u}_\eta-\frac{\chi}{\eta}(\mathbf{u}_\eta-\mathbf{u}_s).
\label{eq_ns2deta}
\end{equation}
\begin{equation}
\nabla\cdot\mathbf{u}_\eta = 0,
\end{equation}
where
the velocity in the solid region is 
$\mathbf{u}_s=(0, u_{s\theta}(r))$ and 
\begin{equation}
u_{s\theta}(r)=\left\{\begin{array}{lll}
        r \omega_1 & \mathrm{in} & \Omega_{s1} \\
\displaystyle
        0          & \mathrm{in} & \Omega_f \\
        r \omega_2 & \mathrm{in} & \Omega_{s2}
        \end{array}
        \right..
\label{eq_us2d}
\end{equation}
For the 2D problem, the original mask function is 
\begin{equation}
\chi(\mathbf{x})=\left\{\begin{array}{lll}
        0 & \mathrm{in} & \Omega_f \\
        1 & \mathrm{in} & \Omega_{s}
        \end{array}
        \right.,
\label{eq_mask2d}
\end{equation}
and the shifted mask function is
\begin{equation}
\chi(\mathbf{x})=\left\{\begin{array}{lll}
        0 & \mathrm{in} & \Omega'_f \\
        1 & \mathrm{in} & \Omega'_{s}
        \end{array}
        \right.,
\label{eq_mask2d2}
\end{equation}
where $\Omega_s = \Omega_{s1} + \Omega_{s2}$ and $\Omega'_s = \Omega'_{s1} + \Omega'_{s2}$.
For the shifted mask function, we define $\Omega'_f=\{ (x, y)\ |\ R_{1\mathrm{VP}}  \le r \le R_{2\mathrm{VP}}\}$, $\Omega'_{s1} = \{ (x, y)\ |\ 0 \le r < R_{1\mathrm{VP}}\}$, and $\Omega'_{s2} = \{ (x, y)\ |\ R_{2\mathrm{VP}} < r \ \mathrm{in} \ \Omega \}$, where $R_{1\mathrm{VP}}=R_1 + \sqrt{\nu\eta}$, $R_{2\mathrm{VP}}=R_2 - \sqrt{\nu\eta}$.
Fig.~\ref{fig_mask3} shows the mask functions near the inner cylinder at various angles.
Since the interfaces between $\Omega_s$ and $\Omega_f$ ($\Omega'_s$ and $\Omega'_f$) are circular, the distribution of the grid points around the interfaces depends on the angle.

\subsection{Numerical setups}
The Fourier pseudo-spectral method and the fourth-order Runge-Kutta method
are used for spatial and time discretization, respectively.
The velocity and pressure are expressed as Fourier series.
The advection and penalization terms are calculated in the physical space,
while the Poisson equation for pressure is solved and the time integration is performed in the Fourier space.
The 2/3 rule is adopted for dealiasing.

The mode number $N$ is varied from 256 to 4096.
The time step $\Delta t$ is $10^{-4}$ or $10^{-5}$
depending on $N$ to meet the numerical stability condition.
In this study $\nu$ is fixed to $10^{-2}$, and $\eta$ is set to $10^{-2}$, $10^{-3}$, and $10^{-4}$.
The parameters for the 2D problem is set to $R = \pi$, $R_1 = 0.4\pi$, $R_2=0.8\pi$, $\omega_1 = 1$, and  $\omega_2 = 0$.

\subsection{Error analysis and validation\label{sec4_2}}

In this section, we examine the characteristics of the total error for the 2D problem, and demonstrate the general applicability of our new approach.

In this section the total error $\delta_\mathrm{tot}$ 
is defined by
\begin{equation}
 \delta_{\mathrm{tot}} \equiv \sqrt{\frac{\int_{\Omega_f} | u_{\theta\eta}(t, \mathbf{x}) - u_{\theta \ \mathrm{exact}} (t, \mathbf{x}) |^{2}d\mathbf{x}}{\int_{\Omega_f}\  d\mathbf{x}}},
\label{eq_errtot3}
\end{equation}
where $u_{\theta\eta}$ is the numerical solution of Eq.~(\ref{eq_ns2deta}), and $u_{\theta \ \mathrm{exact}}$ is the exact solution (\ref{eq_sol_ns2d}). 
For the shifted mask function, $\delta'_\mathrm{tot}$ is defined by Eq.(\ref{eq_errtot3}) replacing $\Omega_f$ by $\Omega'_f$.

\begin{figure}
\begin{center}
\includegraphics[width=7cm]{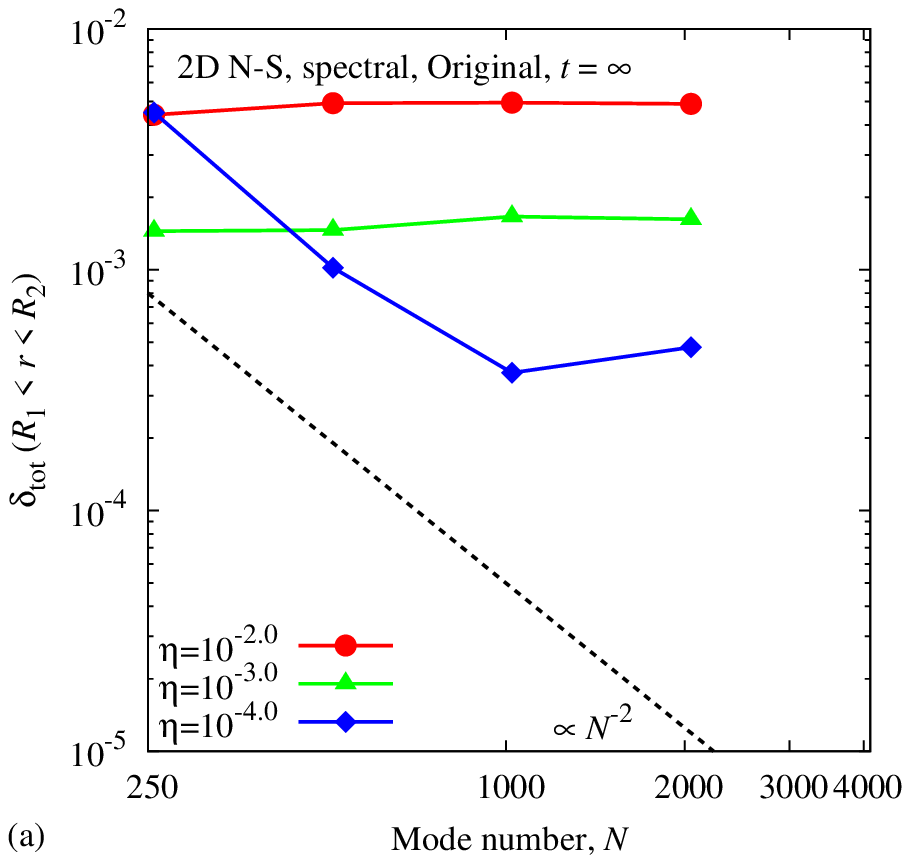}
\includegraphics[width=7cm]{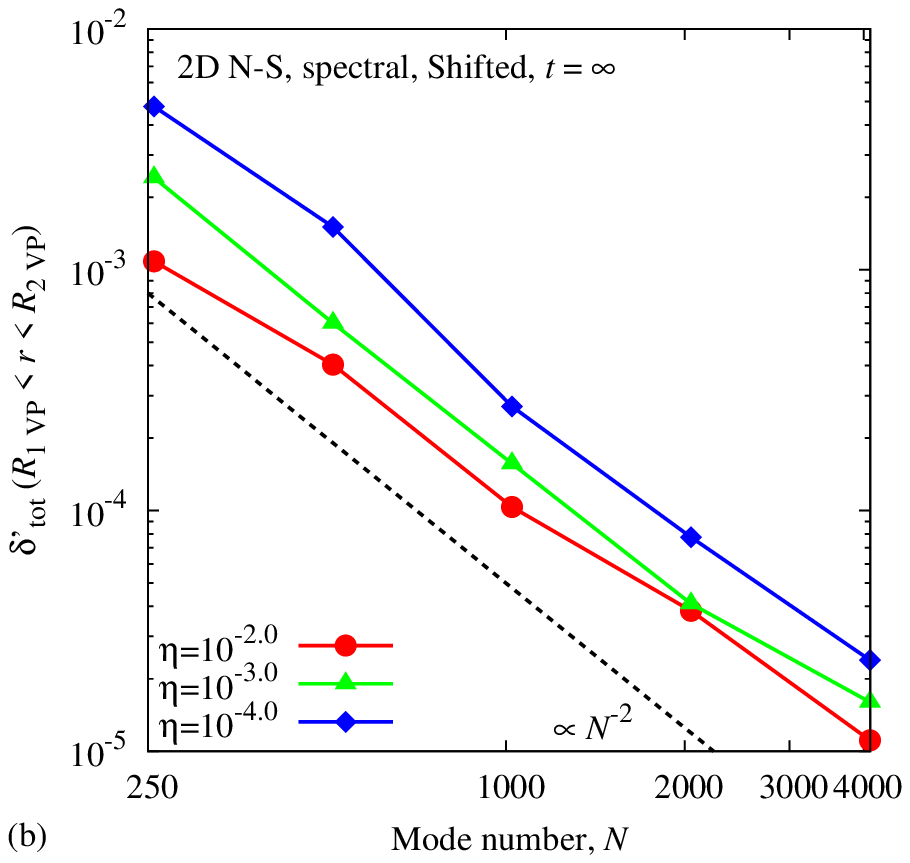}
\end{center}
\caption{
The characteristics of $\delta_\mathrm{tot}$ as a function of $N$ for the 2D N-S equations: (a) the original mask functions, (b) the shifted mask functions.
\label{fig_errns2d}}
\end{figure}

The total error as a function of $N$ for the original and shifted mask functions are shown in Fig.~\ref{fig_errns2d}.
The total error decreases as $N$ increases showing the second-order accuracy for the shifted mask functions,
while it converges to a constant value for large $N$ for the original mask functions.
Moreover, the error is much more reduced for large $\eta$.
It is the same feature as that for the 1D diffusion and Burgers' equations (Figs.~\ref{fig_errdiff1dB} and \ref{fig_errburg1dB}).
This result shows that our new method for error reduction is also effective for the 2D N-S equations.

\begin{figure}
\begin{center}
\includegraphics[width=7cm]{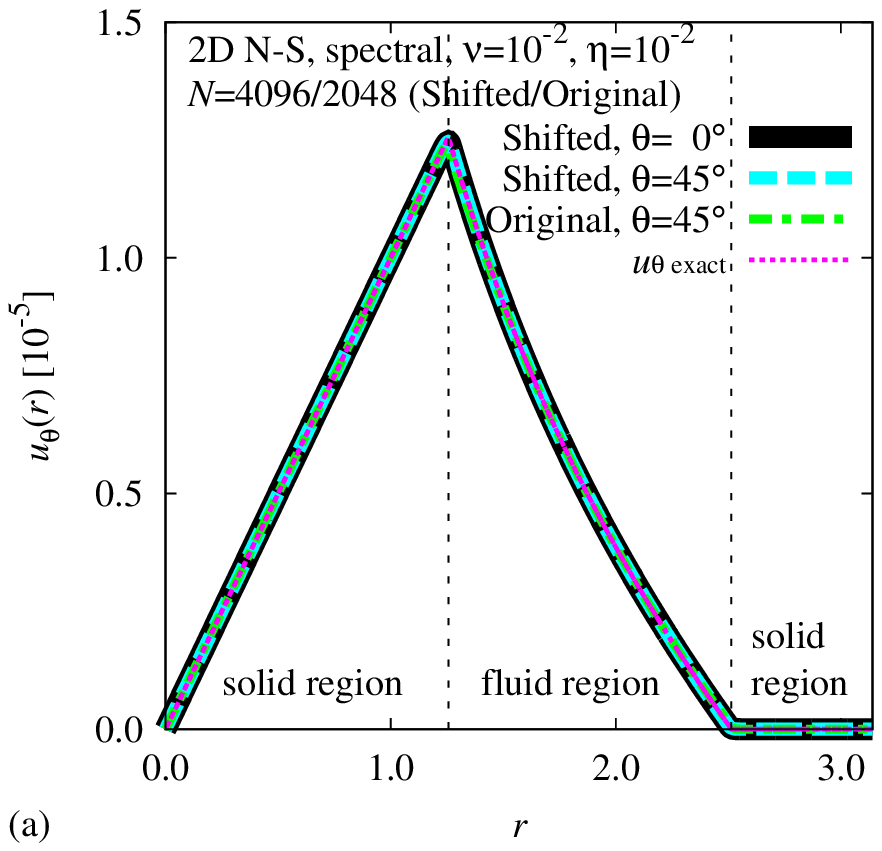}
\includegraphics[width=7cm]{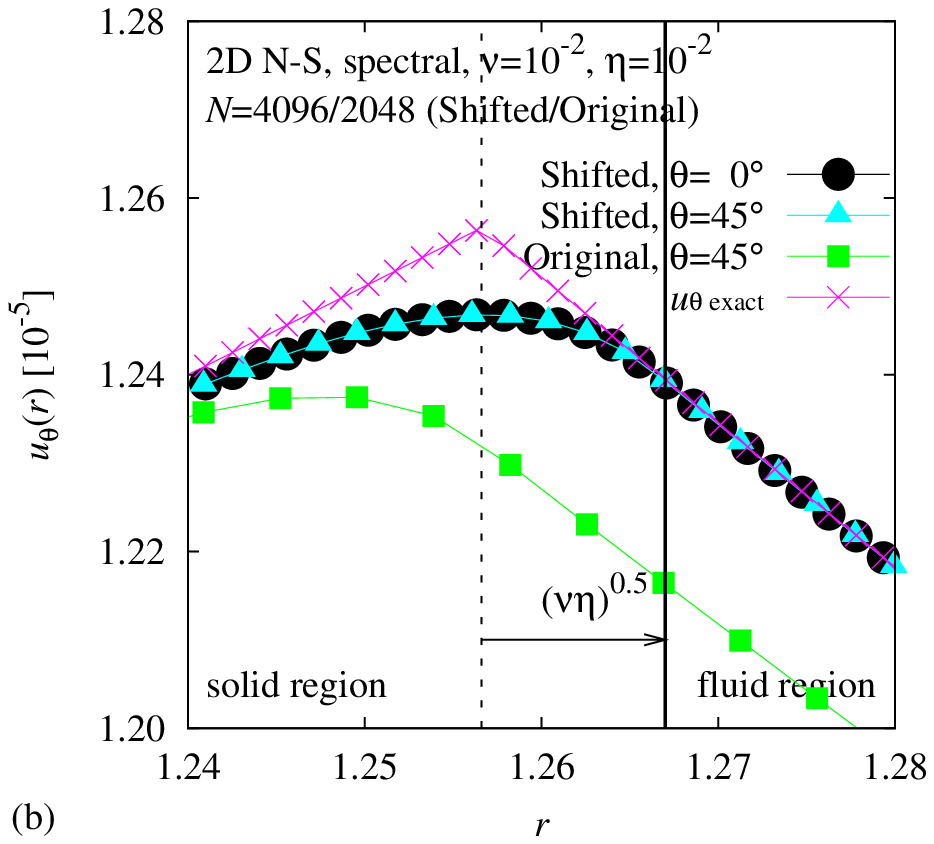}
\includegraphics[width=7cm]{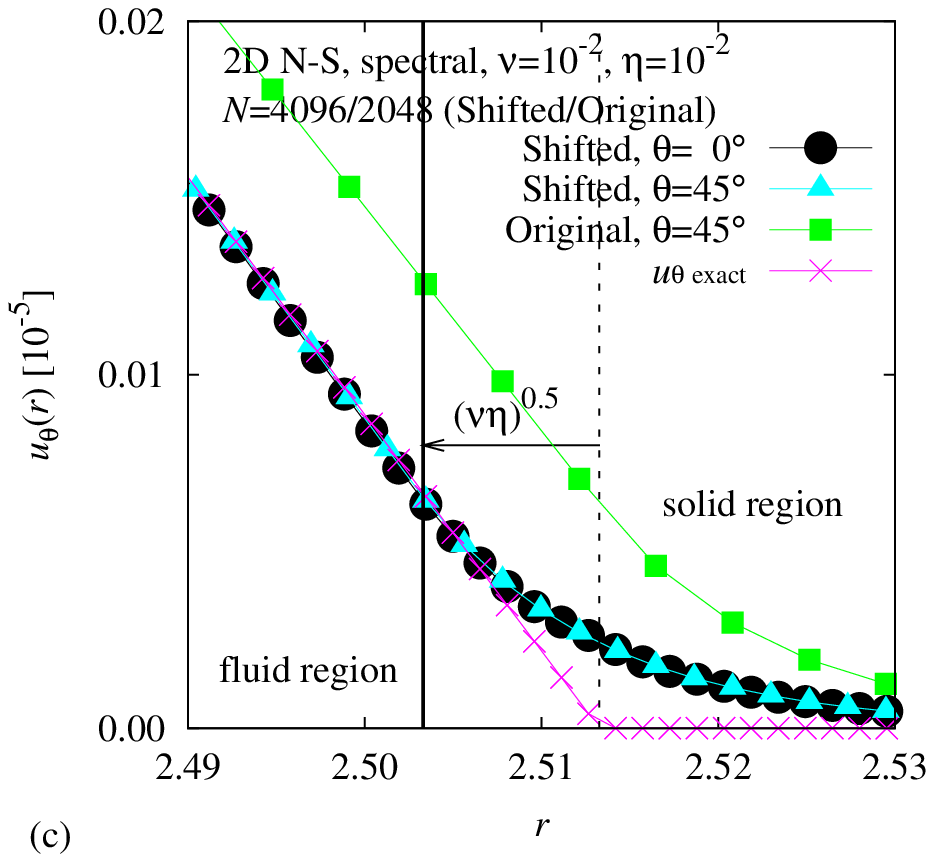}
\end{center}
\caption{
The comparison of solutions for 2D N-S equations. The radial distributions of azimuthal velocity are shown (a) in the whole region, (b) near the inner cylinder, and (c) near the outer cylinder.
\label{fig_velns2d}}
\end{figure}

The solutions in the whole region, near the inner boundary, and near the outer boundary are shown in Fig.~\ref{fig_velns2d}.
The Taylor-Couette flow is well resolved by the pseudo-spectral method (Fig.~\ref{fig_velns2d}a).
Looking at the vicinity of the inner and outer cylinders, however, we observe some differences between the numerical and the exact solutions (Figs.~\ref{fig_velns2d}b and c).
The results obtained by the original type mask function are smaller/larger than the exact solution near the inner/outer boundary.
By shifting the mask function toward the fluid region by $\sqrt{\nu\eta}$,
the numerical solution agrees well with the exact solution in the fluid region except for the immediate vicinity of the cylinders.
These features are exactly the same as those of the 1D diffusion and Burgers' equation.
In addition, the radial distributions for $\theta=0^\circ$ and $\theta=45^\circ$ are in good agreement for $N > 1000$.
Thus, 
if the grid resolution is high enough to capture the boundary layer,
the effect of error reduction is independent of the angle for large $N$,
although the grid system is not spherically symmetric.

\section{Conclusions}

We have investigated the error of the volume penalization method, and have proposed a new method for reducing the error due to the penalization term.
Our findings are summarized below.

First, we found that the mask function Type B, for which the boundary is located at the midpoint of the grid points where the mask function jumps from $0$ to $1$, makes the numerical penalized solutions converge to the exact penalized solutions with highest order.

Next, we modified the mask function in order to reduce 
the numerical error in the volume penalization method.
Our new idea is to shift the boundary between fluid and solid regions of the mask function toward the fluid region by $\sqrt{\nu\eta}$.
The modified mask function makes the total error, 
which is defined as the difference between the numerical solution and the exact solution,
decrease as $N$ increases.
If
the leaking area, 
whose size is $\sqrt{\nu\eta}$,
in the solid region should be adequately resolved for given $N$,
this technique is effective for comparatively large $\eta$.
Thus it is useful when the explicit method is used for time integration 
because of the condition for numerical stability $\Delta t < C \eta$.

Then, we performed numerical simulations of the one-dimensional nonlinear Burgers' equation and the two-dimensional Navier-Stokes equations to confirm the applicability of the present method.
The results showed that modifying the mask function is also effective for the one-dimensional nonlinear problem and the two-dimensional incompressible flow problem of the Taylor-Couette flow between two co-axial cylinders even if circular solid boundaries are immersed in the Cartesian grids.
Therefore, it would be valid for various governing equations, spatial discretization methods, and multi-dimensional problems.

There are several conditions for permeability $\eta$, some of which are mentioned above:
(i) time resolution: response time $\eta$ should be smaller than the smallest time scale which should be resolved;
(ii) spatial resolution: the ``surface thickness'' $\sqrt{\nu\eta}$ should be smaller than the smallest length scale which should be resolved;
(iii) resolution at the boundaries: the grid spacing should be smaller than  $\sqrt{\nu\eta}$ to resolve the surface ``layer'' at the boundaries;
(iv) numerical stability for explicit time integration: $\Delta t < C \eta$.
The conditions (i) and (ii) are necessary,
while (iii) is optional and (iv) is irrelevant when an implicit method is used for time integration.

In this study, we did not discuss the applicability of our error reduction method to the non-uniform grids and moving or deforming solid boundaries.
Furthermore, we did not deal with continuous mask functions\cite{kolomenskiy09}.
These problems are important for development of the volume penalization method and will be investigated as future works.

\section*{Acknowledgments}
Numerical calculations were performed on the Altix UV1000 at the Institute of Fluid Science, Tohoku University.
This study was partially supported by the academic research grant by Maekawa Houonkai in 2011.


\clearpage 

\end{document}